\documentclass[nohyperref]{article}

\usepackage{microtype}
\usepackage{graphicx}
\usepackage{subfigure}
\usepackage{booktabs} 
\usepackage[hyphens]{url}
\usepackage{hyperref}

\usepackage[accepted]{icml2023}

\usepackage{amsmath}
\usepackage{amssymb}
\usepackage{mathtools}
\usepackage{amsthm}

\usepackage[capitalize,noabbrev]{cleveref}

\usepackage{smile}

\theoremstyle{plain}
\newtheorem{theorem}{Theorem}[section]
\newtheorem{proposition}[theorem]{Proposition}
\newtheorem{lemma}[theorem]{Lemma}

\theoremstyle{definition}
\newtheorem{definition}[theorem]{Definition}

\theoremstyle{remark}

\theoremstyle{example}
\newtheorem{example}[theorem]{Example}

\usepackage[textsize=tiny]{todonotes}

\icmltitlerunning{Ranking Losses of CTR Prediction for Welfare Maximization}

\begin{document}

\twocolumn[
\icmltitle{Pairwise Ranking Losses of Click-Through Rates Prediction for Welfare Maximization in Ad Auctions}



\icmlsetsymbol{equal}{*}

\begin{icmlauthorlist}
\icmlauthor{Boxiang Lyu}{booth}
\icmlauthor{Zhe Feng}{google}
\icmlauthor{Zachary Robertson}{stanford}
\icmlauthor{Sanmi Koyejo}{google,stanford}
\end{icmlauthorlist}

\icmlaffiliation{booth}{The University of Chicago Booth School of Business, Chicago, IL USA}
\icmlaffiliation{google}{Google Research, Mountain View, CA USA}
\icmlaffiliation{stanford}{Department of Computer Science, Stanford University, Stanford, CA USA}

\icmlcorrespondingauthor{Boxiang Lyu}{blyu@chicagobooth.edu}

\icmlkeywords{Machine Learning, ICML}

\vskip 0.3in
]



\printAffiliationsAndNotice{}  
\begin{abstract}
We study the design of loss functions for click-through rates (CTR) to optimize (social) welfare in advertising auctions. Existing works either only focus on CTR predictions without consideration of business objectives (e.g., welfare) in auctions or assume that the distribution over the participants' expected cost-per-impression (eCPM) is known a priori, then use various additional assumptions on the parametric form of the distribution to derive loss functions for predicting CTRs.
In this work, we bring back the welfare objectives of ad auctions into CTR predictions and propose a novel weighted rankloss to train the CTR model. Compared to existing literature, our approach provides a provable guarantee on welfare but without assumptions on the eCPMs' distribution while also avoiding the intractability of naively applying existing learning-to-rank methods. Further, we propose a theoretically justifiable technique for calibrating the losses using labels generated from a teacher network, only assuming that the teacher network has bounded $\ell_2$ generalization error. Finally, we demonstrate the advantages of the proposed loss on synthetic and real-world data.

\end{abstract}

\section{Introduction}
\label{sec:intro}
Global online advertising spending is expected to exceed \$700 billion in 2023~\citep{digital}. At the core of online advertising are advertising (ad) auctions, held billions of times per day, to determine which advertisers get the opportunity to show ads~\citep{jeunen2022amazon}. A critical component of these auctions is predicting the click-through rates (CTR)~\citep{yang2022click}. Typically, advertisers submit cost-per-click (CPC) bids, i.e., report how much they are willing to pay if a user clicks. The CTR is the probability that a user clicks the ad when the ad is shown. Combined with the cost-per-click bid, the platform can then calculate the value of \emph{showing} the ad, usually called the cost-per-impression (eCPM). As the CTR needs to be learned, the platform instead uses the predicted click-through rates (pCTRs) to convert the submitted CPC bids to predicted eCPM bids, which then determine the auctions' outcomes.

Due to the importance of predicting the CTRs, a wealth of related literature exists, and we refer interested reader to~\citet{choi2020online,yang2022click} for thorough reviews of these advances. Of these works, the majority focus on the various neural network architectures designed for the task, such as DeepFM~\citep{guo2017deepfm}, Deep \& Cross Network (DCN)~\citep{wang2017deep}, MaskNet~\citep{wang2021masknet}, among many others. These works propose novel neural network architectures but train these networks using off-the-shelf classification losses with no guarantees on the actual economic performance of the ad auctions, creating a discrepancy between the upstream model training for CTR prediction and downstream model evaluation.

Some works aim to ameliorate these discrepancies by using business objectives such as social welfare (or welfare for short) to motivate the design of loss functions~\citep{chapelle2015offline, vasile2017cost, hummel2017loss}. However, these works either lack reproducible experiments on publicly available real-world benchmarks~\citep{hummel2017loss}, or depend on ad-hoc heuristics with insufficient theoretical guarantees~\citep{vasile2017cost}. Moreover, many of these works suffer from an unrealistic assumption that bidders submit eCPM bids and the eCPM of the highest competing bid follows a known and fixed distribution. However, in real life, some ad auctions at industry leaders such as Amazon, Meta, and Google only accept CPC bids~\citep{amazonadshow, metabusinesshelpcenterad, googleadshelphow}, and adjustments to the CTR prediction model changes the distribution of competing bids' eCPM.

We avoid the pitfalls of existing works by limiting assumptions about the eCPMs' distribution. Since various types of ad auctions with drastically different revenue functions are widely deployed, ranging from Generalized Second Price~\citep{edelman2007internet} to Vickrey-Clarke-Groves~\citep{varian2014vcg}, and first price auction~\citep{conitzer2022pacing}, we focus on maximizing the welfare achieved by these auctions, which measures the efficiency of the ad auction in terms of showing the most valuable ads. 

{\noindent \textbf{Our Contributions.}} We list our contributions below.
\vspace{-1em}
\begin{itemize}
\itemsep0em 
    \item We propose a learning-to-rank loss with welfare guarantees by drawing a previously underutilized connection between welfare maximization and ranking.
    \item We propose two surrogate losses that are easy to optimize and theoretically justifiable.
    \item Inspired by student-teacher learning~\citep{hinton2015distilling}, we construct an approximately calibrated, easy-to-optimize surrogate, whose theoretical guarantees only depend on the $\ell_2$-generalization bound of the teacher network.
    \item We demonstrate the benefits of the proposed losses on both simulated data and the Criteo Display Advertising Challenge dataset\footnote{\url{https://www.kaggle.com/c/criteo-display-ad-challenge}}, arguably the most popular benchmark for CTR prediction in ad auctions.
\end{itemize}

\subsection{Related Works}
In this section, we divide the related works into three main categories: applied research in CTR prediction, theoretical analysis of ad auctions, and methods in learning-to-rank.

{\noindent \textbf{Applied Research in CTR Prediction.}} There is an abundance of application oriented literature on CTR prediction~\citep{mcmahan2013ad,chen2016deep,cheng2016wide,zhang2016deep,qu2016product,juan2017field,lian2018xdeepfm,zhou2018deep,zhou2019deep,wang2021masknet,pi2019practice,pan2018fieldweighted,li2020interpretable,chapelle2015offline}, and we refer interested readers to~\citet{yang2022click} for a detailed survey. Two works with well-documented performance on the Criteo dataset are~\citet{guo2017deepfm} and~\citet{wang2017deep}. Particularly,~\citet{guo2017deepfm} proposes DeepFM, short for deep factorization machines, which combines deep learning with factorization machines.~\citet{wang2017deep} is similar, where the proposed Deep Cross Network model combines deep neural networks with cross features. These works focus on the development of neural network architectures and use classification losses with little to no theoretical guarantees. Our work is orthogonal to and complements this line of literature by proposing easy-to-optimize loss functions rooted in economic intuition with provable guarantees on economic performance. 

A well-known technique in knowledge distillation is student-teacher learning~\citep{hinton2015distilling}, where a smaller network is used to approximate the predictions of a larger one. Recently some attempts have been made at applying the technique in CTR prediction~\citep{zhu2020ensembled} and, as we demonstrate in this manuscript, the technique can even benefit the design of welfare-inspired loss functions, in addition to reducing the computation and memory requirements of the teacher network itself.

Among this line of work, two papers are closer to ours in spirit.~\citet{chapelle2015offline} studies the design of CTR evaluation metrics that approximate the bidders' expected utility.
Similarly, \citet{vasile2017cost} uses the utility that the bidder derives from the auction to design a suitable loss function that the bidder should use for CTR prediction. While both works provide empirical justifications for the proposed losses, they only provide heuristic arguments when designing the loss functions themselves and include no theoretical guarantees on the generalization or calibration of the losses. Moreover, they both rely on the assumption that the distribution of the highest competing bid's eCPM is fixed and known a priori.

{\noindent \textbf{Theoretical Analysis of Ad Auctions.}} Many works study the theoretical properties of ad auctions~\citep{fu2012ad,edelman2010optimal,gatti2012truthful,aggarwal2006truthful,varian2009online,dughmi2013constrained,bergemann2022calibrated,lucier2012revenue}, and~\citet{choi2020online} offers a detailed survey of a collection of recent advances in the analysis of ad auctions.

\citet{hummel2017loss} is the most relevant work to ours, as it studies the design of loss functions in ad auctions from the seller's perspective, offering new insights on how to design losses for either welfare maximization or revenue maximization. However, the real-world experiments in the paper rely on proprietary data, and the claims are not verified on widely available benchmarks. Moreover, it again relies heavily on the assumption that the distribution of the highest competing bid's eCPM is known beforehand, which can be unrealistic in practice.

{\noindent \textbf{Learning-to-Rank.}} Our work draws inspiration from a line of research on learning-to-rank~\citep{burges2005learning,burges2006learning,cortes2010learning,burges2010ranknet,wang2018lambdaloss}, which incorporates information retrieval performance metrics such as Normalized Discounted Cumulative Gain into the design of the loss functions, resembling our works. However, as we show in Section~\ref{sec:failure_of_learning_to_rank}, these works do not directly apply to the welfare maximization setting. Moreover, to the best of our knowledge, these works have not been examined in the context of welfare maximization in ad auctions.


\section{Models and Preliminaries}
We begin with a multi-slot ad auction~\cite{edelman2007internet, Varian07} where each ad is associated a cost-per-click (CPC) bid. Let $K$ denote the number of the slots and each slot, indexed by $k$, is associated with a position multiplier $\alpha_k$. Without loss of generality assume that $\alpha_1 = 1$ and the weights are decreasing in $k$, namely $\alpha_1 \geq \ldots \geq \alpha_K$. Assume there are $n \geq K$ ads participating in the auction where each ad has a feature vector $x_i \in \RR^d$ and CPC bid $b_i$. There exists a function $p^*: \RR^d \to [0,1]$ such that the CTR of the ad at slot $k$ is $\alpha_k p_i$, where we let $p_i = p^*(x_i)$ for convenience. The ad's CTR is affected by both the slot it is assigned to and the ad's features. Intuitively, $p_i$ is the ad's base CTR if it were assigned to the first slot, and is scaled according to $\alpha_k$ for any slot $k$.

Throughout this paper, we assume that the position multipliers are known, and we focus only on learning $p^*$, i.e., the ad's CTR if it were assigned the first slot. Learning a position-based CTR prediction model requires additional assumptions to model the user's click behavior and is outside of our scope, which focuses on welfare maximization instead. Indeed, we will show it is without loss of generality to focus on single-slot auctions to maximize welfare, which is equivalent to learning the base CTR when shown in the first slot (Proposition~\ref{prop:welfare_maximization_reduction}).

More concretely let $\cH \subseteq \{f: \cX \to [0,1]\}$ denote the hypothesis space and assume that $p^*$ is realizable, i.e. $p^*(\cdot) \in \cH$. Conditioned on a set of $n$ ads $\{(b_i, x_i)\}_{i = 1}^n$, let $f(\cdot)$ denote an arbitrary function that the seller uses to predict the CTRs. The function $f$, combined with the submitted bid $b_i$ and the observed context $x_i$, yields the predicted eCPMs $b_if(x_i)$ for all $i \in [n]$. The seller then awards the first slot to the bidder with the highest predicted eCPM, the second slot to the bidder with the second, and so forth, achieving a welfare of
\begin{align*}
    \mathrm{Welfare}_f(\{(b_i, x_i)\}_{i = 1}^n) &= \sum_{k = 1}^K b_{\pi_f(k)}p_{\pi_f(k)},
\end{align*}
where for any function $f$, $\pi_f(k)$ returns the index of the ad with the $k$-th highest predicted eCPM. The welfare maximization problem is then
\begin{equation}
    \label{eq:defn_welfare_maximization_problem}
    \begin{aligned}
        \max_{f \in \cH} \sum_{k = 1}^K b_{\pi_f(k)}p_{\pi_f(k)}.
    \end{aligned}
\end{equation}
As we will prove, a solution to the problem is $f = p^*$.
For convenience, we let $\pi_*(\cdot) = \pi_{p^*}(\cdot)$, $\mathrm{Welfare}_*(\{(b_i, x_i)\}_{i = 1}^n) = \mathrm{Welfare}_{p^*}(\{(b_i, x_i)\}_{i = 1}^n)$, and assume there are no ties in $b_if(x_i)$ or $b_ip_i$.

To better illustrate welfare and advertisement auction, we include an specific instance of ad auction in the following example.

\begin{example}
    \label{example}
    Let $\texttt{ad}_1, \texttt{ad}_2, \texttt{ad}_3$ denote three different advertisements, where $\texttt{ad}_1$'s CTR is 0.1 and CPC bid 10, $\texttt{ad}_2$'s CTR 0.4 and CPC bid 2, and $\texttt{ad}_3$'s CTR 0.9 and CPC bid 0.5. Suppose there two advertisement slots where the first slot has multiplier $\alpha_1 = 1$ and the second $\alpha_2 = 0.9$. Assigning the first slot to $\texttt{ad}_1$ and the second to $\texttt{ad}_2$ maximizes welfare, and the maximum welfare is $ 1 + 0.8 \times 0.9 = 1.62$. Knowing the ads' exact CTR helps us achieve this maximum welfare.
\end{example}

\subsection{Welfare Maximization and Ranking}
We first show that we lose no generality by restricting our focus to single-slot ad (e.g., the first slot) auctions. 
\begin{proposition}[Reduction to Single-slot Setting]
    \label{prop:welfare_maximization_reduction}
    The function $f$ maximizes welfare in a $K$-slot auction only if it maximizes welfare in single-slot ad auctions held over subsets of the participating ads. Moreover, the ground-truth CTR function $p^*$ maximizes welfare.
\end{proposition}

Detailed proof of the proposition is deferred to Appendix~\ref{sec:proof_prop_welfare_maximization_reduction}. Consider the setting in Example~\ref{example}, for instance. Only considering the welfare objective, note that we can auction off the two ad slots one by one, where $\texttt{ad}_1, \texttt{ad}_2, \texttt{ad}_3$ participates in the auction for the first slot and $\texttt{ad}_2, \texttt{ad}_3$ participates in that for the second. In this setting, if we know the ads' ground-truth CTR, then $\texttt{ad}_1$ wins the first slot and $\texttt{ad}_2$ wins the second, achieving the maximum welfare. 

By Proposition~\ref{prop:welfare_maximization_reduction}, we can see that welfare maximization in multi-slot ad auctions is no harder than welfare maximization in single-slot ad auctions, and this relies on the fact that the position multipliers are independent of advertisers. For the rest of the paper, we then without loss of generality focus only on single-slot auctions.

As welfare is maximized by the ground-truth CTR function, a common approach is to treat the problem as a classification problem, using $y_i$ as feedback for learning $p^*$~\citep{vasile2017cost, hummel2017loss}. However, as noted in Section~\ref{sec:intro}, this approach can suffer from a mismatch between the loss function and the business metric (in our case, welfare). 

We notice that welfare maximization can be reduced to a learning-to-rank problem instead. Let $i^* = \pi_*(1)$ be the index of the ad with the highest ground-truth eCPM and $j^* = \pi_f(1)$ be that of the ad with the highest predicted eCPM. We note that
\begin{equation}
    \label{eq:pair_welfare_loss}
    \begin{split}
        &\mathrm{Welfare}_*(\{(b_i, x_i)\}_{i = 1}^n) - \mathrm{Welfare}_f(\{(b_i, x_i)\}_{i = 1}^n)\\
        &\qquad = \sum_{i = 1}^n \sum_{j = 1}^n ((b_ip_i - b_jp_j)\ind\{i = i^*\}\ind\{j = j^*\}\\
        &\hspace{11em}\times\ind\{b_if(x_i) \leq b_jf(x_j)\}).
    \end{split}
\end{equation}
We defer the detailed derivation of~\eqref{eq:pair_welfare_loss} to Appendix~\ref{sec:proof_thm_lower_bounds_welfare}. Since $b_{i^*}p_{i^*}$ yields the highest ground-truth eCPM, welfare is maximized if and only if $j^* = i^*$. Consequently, as long as $f$ correctly \emph{ranks} each pair of observations according to their ground-truth eCPM, it also correctly identifies the ad with the highest ground-truth eCPM and maximizes welfare. The reduction to ranking generalizes to multi-slot auctions, and we defer a formal statement to Lemma~\ref{lemma:welfare_maximization_ranking} in the appendix.

The same intuition is illustrated by Example~\ref{example}: as long as we can rank the three ads according to their ground-truth eCPM ($\texttt{ad}_1 > \texttt{ad}_2 > \texttt{ad}_3$), then we can maximize the auction's welfare.

To summarize, we must rank the ads according to their ground-truth eCPM using a suitable CTR prediction function to maximize welfare. An approach that follows this observation is to learn a CTR prediction rule to rank the ads, leading to the proposed ranking-inspired losses.


\section{Ranking-Inspired Loss Functions for Welfare Maximization}
Let $\cD = \{(b_i, x_i), y_i\}_{i = 1}^n$ be a batch of $n$ ads participating in one round of an ad auction, where $y_i \sim \mathrm{Ber}(p_i)$ indicates whether the ad has been clicked or not. 
We then call $b_ip_i$ the ad's ground-truth eCPM and $b_iy_i$ its empirical eCPM.
 Consider the following pairwise loss function (which we propose the seller minimize):
\begin{equation}
    \label{eq:indicator_pairwise_surrogate}
    \ell(f; \cD) = \sum_{i = 1}^n\sum_{j = 1}^n\left({b_iy_i} - {b_jy_j}\right)\ind\{b_if(x_i) \leq b_jf(x_j)\}.
\end{equation}
Let $\cR(f; \cD) = \EE_{\{y_i\}_{i = 1}^n}[\ell(f; \cD)]$ denote the conditional risk induced by the loss function $\ell$. Recalling that $y_i \sim \textrm{Bernoulli}(p_i)$, we know
\begin{equation}
    \label{eq:defn_conditional_risk}
    \cR(f; \cD) = \sum_{i = 1}^n\sum_{j = 1}^n(b_ip_i - b_jp_j)\ind\{b_if(x_i) \leq b_jf(x_j)\}.
\end{equation} 
Observe the similarities between~\eqref{eq:pair_welfare_loss} and~\eqref{eq:defn_conditional_risk}. The conditional risk $\cR(f; \cD)$ can be viewed as a proxy for the welfare suboptimality of $f$, where we replace $\ind\{i = i^*\}\ind\{j = j^*\}$ with 1. While $j^*$ is easy to determine once $f$ is given, we do not the index with the highest ground-truth eCPM. Fortunately, as we show in the following proposition, minimizing $\cR(f; \cD)$ via empirical risk minimization is a reasonable proxy for minimizing welfare suboptimality.

\begin{proposition}
    \label{thm:recovers_correct_ranking}
    For any $\cD$, let $\hat{f}$ be an arbitrary and fixed minimizer of the conditional risk $\cR(f; \cD)$. We then know $\hat{f}$ ranks every pair in the sequence correctly, i.e. $b_ip_i \geq b_jp_j$ if and only if $b_i\hat{f}(x_i) \geq b_j\hat{f}(x_j)$. Moreover, the ground-truth CTR function $p^*(\cdot)$ minimizes the conditional risk $\cR(f; \cD)$ for any $\cD$.
\end{proposition}
Detailed proof for the proposition can be found in Appendix~\ref{sec:proof_thm_recovers_correct_ranking}. 
Proposition~\ref{thm:recovers_correct_ranking} shows that minimizing the conditional risk $\cR(f; \cD)$ is a surrogate for maximizing welfare and minimizing~\eqref{eq:indicator_pairwise_surrogate} is a reasonable choice of loss function. In the following theorem, we make explicit the connection between the conditional risk and welfare. With a slight abuse of notation let $\mathrm{Welfare}_f(\cD), \mathrm{Welfare}_*(\cD)$ denote the welfare achieved by $f$ and the optimal (achievable) welfare, respectively,  when $\{(b_i, x_i)\}_{i = 1}^n$ are given by the dataset $\cD$. We emphasize the conditional risk $\cR(f; \cD)$ can be negative, an important fact to bear in mind in the context of the following theorem.
\begin{theorem}\label{thm:lower_bounds_welfare}
    The following holds for all $f \in \cH$ and $\cD$
    \begin{align*}
        \mathrm{Welfare}_*(\cD) &\leq \mathrm{Welfare}_f(\cD) +\frac{1}{2}\cR(f; \cD)\\
        &\qquad + \frac{1}{4}\sum_{i = 1}^n\sum_{j = 1}^{n} \abr{b_ip_i - b_jp_j}.
    \end{align*}
     Moreover, the bound is tight for any minimizer of $\cR(f; \cD)$ and for all $\cD$
     \[
        \min_{f \in\cH} \cR(f; \cD) = -\frac{1}{2}\sum_{i = 1}^n\sum_{j = 1}^{n} \abr{b_ip_i - b_jp_j}.
     \]
\end{theorem}
See Appendix~\ref{sec:proof_thm_lower_bounds_welfare} for detailed proof. We note that the theorem provides a valid lower bound for all possible $f \in \cH$. More importantly, for any dataset $\cD$, we can show that there is at least one minimizer of $\cR(f; \cD)$ thanks to the realizability assumption, for which $\frac{1}{2}\cR(f; \cD) + \frac{1}{4}\sum_{i = 1}^n\sum_{j = 1}^{n} \abr{b_ip_i - b_jp_j} = 0.$ Crucially, the theorem implies that minimizing the conditional risk on any dataset $\cD$ maximizes welfare, further justifying the use of $\ell(f; \cD).$

While we have shown minimizing $\cR(f; \cD)$ suffices for welfare maximization, recovering the ground-truth CTR function $p^*(\cdot)$ remains crucial for real-world ad auctions. For instance, revenue in generalized second price auctions depends on the pCTRs themselves, and functions that correctly rank the ads do not necessarily lead to high revenue. Fortunately, by adding a calibrated classification loss to $\ell(f; \cD)$, we can ensure that $p^*(\cdot)$ minimizes the (unconditional) risk. Particularly, we have the following proposition.
\begin{proposition}
    \label{prop:calibrated_combo_loss}
    Let $h(f; \cD)$ denote an arbitrary loss function such that $p^*$ is the unique minimizer of $\EE_{\cD}[h(f; \cD)]$.
    For any constant $\lambda > 0$,  $p^*$ is the unique minimizer of $\EE_{\cD}[\ell(f; \cD) + \lambda h(f;\cD)]$.
\end{proposition}
See Appendix~\ref{sec:proof_prop_calibrated_combo_loss} for detailed proof.
We note that logistic loss and mean squared error are both valid choices for $h(f; \cD)$ in Proposition~\ref{prop:calibrated_combo_loss}.

\subsection{Easy-to-Optimize Surrogates}
While $\ell(f; \cD)$ is attractive as it is closely related to the welfare, the function itself is nondifferentiable and cannot be efficiently optimized using first-order methods (e.g., SGD) due to the indicator variables. We thus propose two differentiable surrogates with provable performance guarantees
\begin{equation}
\label{eq:defn_log_surrogate}
\begin{split}
\ell^{\log}_{\sigma}(f; \cD)
& =\sum_{i = 1}^n\sum_{j = 1}^n (b_iy_i - b_jy_j) \\
&\qquad \times \log(1 + \exp(-\sigma(b_if(x_i) - b_jf(x_j)))),
\end{split}
\end{equation}
and
\begin{equation}
    \label{eq:defn_hinge_surrogate}
    \begin{split}
        \ell^{\rm hinge}_{\sigma}(f; \cD)&= \sum_{i = 1}^n\sum_{j = 1}^n (b_iy_i - b_jy_j)\\
        &\qquad \times (-\sigma(b_if(x_i) - b_jf(x_j)))_{+}.
    \end{split}
\end{equation}

For~\eqref{eq:defn_log_surrogate}, we replace the indicators in $\ell(f; \cD)$ with the log logistic function $-\log(1 + \exp(-\sigma (b_if(x_i) - b_jf(x_j))))$. Similarly,~\eqref{eq:defn_hinge_surrogate} acts as a surrogate to $\ell(f; \cD)$ with the indicator replaced by $(-\sigma(b_if(x_i) - b_jf(x_j)))_{+}$ instead, where for any $a \in \RR$ we let $(a)_+=\max(0,a)$. While the function $(\cdot)_+$ itself is not differentiable at $x = 0$, it is differentiable almost everywhere and can be easily optimized using its subderivative.

\begin{figure}[ht]
    \centering
    \includegraphics[width=0.9\linewidth]{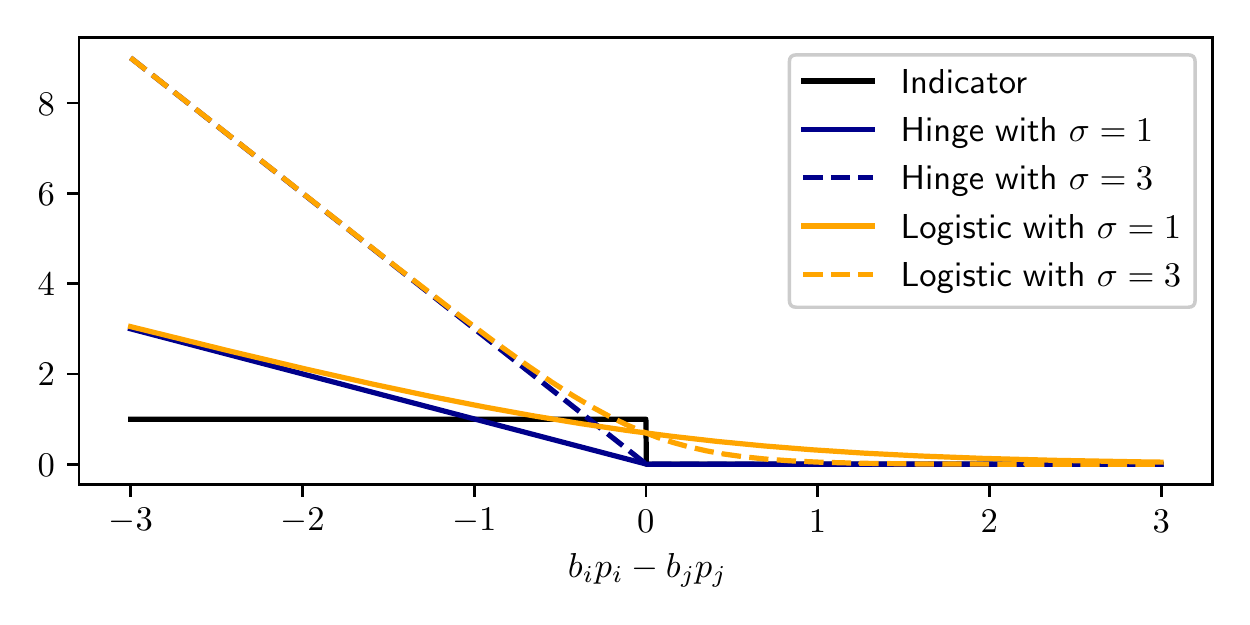}
    \caption{Visualization of the surrogates to $\ind\{b_i f(x_i) \leq b_jf(x_j)\}$ for different values of $\sigma$ as functions of $b_i f(x_i) - b_jf(x_j)$.}
    \label{fig:ind_surrogates}
\end{figure}

For both surrogates, the term $\sigma$ is a manually adjustable parameter controlling how much we penalize small margins between a pair of eCPM, $b_if(x_i) - b_jf(x_j)$. As we can see from Figure~\ref{fig:ind_surrogates}, for pairs of ads whose predicted eCPMs are close to each other, a larger $\sigma$ accentuates the difference between them and leads to a surrogate value close to one. However, as $\sigma$ increases, the surrogate value for ads with large gaps in predicted eCPMs tend to be much larger than one. Adjusting $\sigma$ is then a balancing act between these two kinds of pairs.

Regardless of the choice of surrogate for the indicator function, the surrogate losses themselves remain closely related to~\eqref{eq:pair_welfare_loss}, which we highlight in the following theorems.

\begin{theorem}\label{thm:log_surrogate}\label{thm:hinge_surrogate}
    Assuming all bids are bounded by some $B \in \RR_{>0}$, setting $\sigma = 2/B$ ensures for any $f \in \cH$ and $\cD$
    \begin{align*}
        |\EE_{\{y_i\}_{i = 1}^n}[\ell^{\log}_{\sigma}(f; \cD)] - \cR(f; \cD)| &\leq \Delta,\\
        |\EE_{\{y_i\}_{i = 1}^n}[\ell^{\rm hinge}_{\sigma}(f; \cD)] - \cR(f; \cD)| &\leq \Delta,
    \end{align*}
    where $\Delta = \frac{1}{2}\sum_{i = 1}^n\sum_{j =1}^n|b_ip_i - b_jp_j|$ is a problem-dependent constant.
\end{theorem}

See Appendix~\ref{sec:proof_thm_smooth_surrogate} for detailed proof.
Theorem~\ref{thm:log_surrogate} shows that $\ell_\sigma^{\log}(f; \cD)$ and $\ell_\sigma^{\rm hinge}(f; \cD)$ are closely tied to the original loss $\ell(f; \cD)$. While assuming the CPC bids are bounded implicitly implies the eCPMs' are also bounded, the assumption is mild and does not restrict the parametric form of the eCPMs' distribution. While the surrogates do not exactly match the proposed loss $\ell(f; \cD)$, the gap is due to approximating the indicators in $\ell(f; \cD)$ and cannot be avoided.

\subsection{Failure of Directly Applying Learning-to-Rank}
\label{sec:failure_of_learning_to_rank}
It may be tempting to further exploit the connection between welfare and ranking over predicted eCPMs by applying a learning-to-rank loss function directly on the observed eCPMs $b_iy_i$. As we show below, the approach, unfortunately, fails, as the inclusion of bids makes the empirical observation $\ind\{b_iy_i \geq b_jy_j\}$ a poor estimate of $\ind\{b_ip_i \geq b_jp_j\}$. 

\begin{proposition}\label{prop:direct_learning_to_rank_fails}
    For any $1/2 > \epsilon > 0$, there exists a pair of ads $i$ and $j$ such that
    \[
        \Pr(\ind\{b_i y_i \geq b_j y_j\} = \ind\{b_i p_i \geq b_j p_j\}) = \epsilon,
    \]
    where $(b_i, p_i, y_i)$ and $(b_j, p_j, y_j)$ are the CPC bids, ground-truth CTR, and realized click indicator for the two ads.
\end{proposition}
See Appendix~\ref{sec:proof_prop_direct_learning_to_rank_fails} for detailed proof.
Intuitively, the construction of the counterexample in Proposition~\ref{prop:direct_learning_to_rank_fails} relies on the fact that the ground-truth eCPM of an ad increases as its corresponding CPC bid increases, but the probability that the ad is clicked does not. In other words, for any ad $i$, the probability that $b_iy_i$ is non-zero does not depend on $b_i$ while the ground-truth eCPM does, creating a discrepancy between the ground-truth eCPM and the empirical eCPM. We may then strategically manipulate $b_i$ to construct an example satisfying Proposition~\ref{prop:direct_learning_to_rank_fails}.

Crucially, Proposition~\ref{prop:direct_learning_to_rank_fails} shows that there exist pairs of ads whose empirically observed CPM rankings agree with their ground-truth eCPM rankings with probability arbitrarily close to zero. Unless strong assumptions are made on the distributions of empirically observed CPMs, it is impossible to directly apply off-the-shelf learning-to-rank loss functions for $\ind\{b_iy_i \geq b_jy_j\}$.

On the other hand,~\eqref{eq:indicator_pairwise_surrogate} avoids the pitfall by weighing each entry by $(b_iy_i - b_jy_j).$ When conditioned on any CTR prediction rule $f(\cdot),$ by the linearity of expectation, we can see that the weight is an unbiased estimate of the difference in ground-truth eCPM. The fact that $\ell(f; \cD)$ is linear in each observed eCPM $b_iy_i$ is crucial, as the linearity ensures that the loss function accurately reflects the differences in $b_ip_i$, enabling us to relate the conditional risk to the actual welfare loss and obtain theoretical guarantees without any assumptions on the empirical eCPMs. 

To the best of our knowledge, no existing works on learning-to-rank use loss functions of this form, and our proposed methods are uniquely capable of avoiding the challenge highlighted by Proposition~\ref{prop:direct_learning_to_rank_fails}. While resembling a learning-to-rank loss, \eqref{eq:indicator_pairwise_surrogate} is at its core a loss function that resembles the shape of the welfare objective in an ad auction, ensuring that optimizing the loss is closely related to optimizing welfare.

\section{Replacing $y_i$ with Predictions from the Teacher Model}
A concern for~\eqref{eq:defn_log_surrogate} and~\eqref{eq:defn_hinge_surrogate} is that their variance scales with $b_i^2$, the squared values of the CPC bids. Combined with the noisiness of $y_i$, the resulting loss may be overly noisy. While the issue might be mitigated by properly pre-processing the CPC bids, we propose a theoretically justifiable alternative inspired by student-teacher learning~\citep{hinton2015distilling}. In fact, distillation loss associated with the prediction from the teacher model is widely used in industrial-scale advertising systems \citep{google-pctr}. This technique is shown to be helpful for stabilizing the training and improving the pCTR accuracy of the student model.

The idea is straightforward. Let $\hat{p}(\cdot)$ be a teacher network trained on the same dataset and we replace $y_i$ with $\hat{p}(x_i)$. We show that doing so leads to an empirical loss that is close to $\cR(f; \cD)$, the conditional risk, as long as the teacher network itself is sufficiently accurate. We begin with the following theorem for replacing the $y_i$'s in~\eqref{eq:indicator_pairwise_surrogate}.
\begin{theorem}\label{thm:plug_in_estimate}
    Let $\hat{p}$ be an estimate of $p^*$ such that $\EE_{x}[(\hat{p}(x) - p^*(x))^2] \leq \epsilon.$ Let $\hat{\ell}(f; \cD)$ be~\eqref{eq:indicator_pairwise_surrogate} but with each $y_i$ replaced by $\hat{p}_i = \hat{p}(x_i)$, i.e.,
    \[
        \hat{\ell}(f; \cD) = \sum_{i = 1}^n\sum_{j = 1}^n (b_i\hat{p}_i - b_j\hat{p}_j)\ind\{b_if(x_i) \geq b_jf(x_j)\}.
    \] 
    Assuming all bids are upperbounded by positive constant $B \in \RR_{> 0}$, for any $f \in \cH$ we have
    \[
        \EE_{\cD}[|\hat{\ell}(f; \cD) - \cR(f; \cD)|] \leq (n - 1)nB\sqrt{\epsilon}
    \] 
    where $n$ is the number of ads.
\end{theorem}
See Appendix~\ref{sec:proof_thm_plug_in_estimate} for proof.
As $\hat{\ell}(f; \cD)$ sums over all pairs of ads, the bound necessarily grows in $\cO(n^2)$, and the factor can be removed if we use the average over the pairs instead.

While the teacher network may be used in ad auctions as-is, student networks still offer several benefits in addition to the theoretical guarantee in Theorem~\ref{thm:plug_in_estimate}. First, teacher networks may be costly to deploy, thus student networks offer efficiency benefits from knowledge distillation. Second, the ranking losses may help the student network better differentiate the eCPMs of pairs of ads, leading to higher welfare, as we observe in experiments.

It is also reasonable to suggest directly learn-to-rank with $\ind\{b_i \hat{p}(x_i)\geq b_j \hat{p}(x_j)\}$ as the labels. However, theoretical guarantees for the approach require additional assumptions on the distribution of the gaps between pairs of predicted eCPM, which is not needed for Theorem~\ref{thm:plug_in_estimate}.

Recalling Theorem~\ref{thm:log_surrogate} and Theorem~\ref{thm:hinge_surrogate}, it is not hard to see that replacing $y_i$ with $\hat{p}(x_i)$ in~\eqref{eq:defn_log_surrogate} and~\eqref{eq:defn_hinge_surrogate} lead to losses that are also sufficiently close to $\cR$. 
We instead focus on using the teacher network to improve calibration.

\subsection{Improving Calibration with the Teacher Network}
A drawback shared by~\eqref{eq:defn_log_surrogate} and~\eqref{eq:defn_hinge_surrogate} is that they are not calibrated. While both penalizes pCTR functions for incorrectly ranking pairs of ads, they also reward pCTR functions that overestimate the margin between pairs of ads. As the minimizers of their expected values are not necessarily the ground-truth CTR function, using these losses may have negative consequences on other important metrics such as revenue or area under the curve.
Fortunately, we show that using a teacher network also improves the calibration of the loss function. We propose the following loss function.

\begin{equation}
\label{eq:defn_hinge_surrogate_plug_in}
    \begin{split}
        \hat{\ell}_{\sigma}^{\textrm{hinge}, +}(f; \cD)&= \sum_{i = 1}^n \sum_{j = 1}^n (b_i\hat{p}_i - b_j \hat{p}_j)_+ \\
        &\qquad \times (-\sigma(b_if(x_i) - b_jf(x_j)))_+,
    \end{split}
\end{equation}
Intuitively, $\hat{\ell}_{\sigma}^{\textrm{hinge}, +}(f; \cD)$ no longer punishes $f$ for having a small margin between predicted eCPMs, as long as $f$ ranks the pair the same way $\hat{p}$ does. When the teacher network is sufficiently close to the ground-truth, the loss function eliminates the bias that~\eqref{eq:defn_log_surrogate} and~\eqref{eq:defn_hinge_surrogate} have towards functions with larger margins between pairs. Additionally, compared to directly using $\hat{p}(\cdot)$,~\eqref{eq:defn_hinge_surrogate_plug_in} better reflects the impact that the pCTRs have on welfare and has theoretical guarantees in terms of welfare performance. 

We now present theoretical justification for the approach. 
Recall from~\citet{vasile2017cost} that calibration in ad auctions is defined as follows.
\begin{definition}[Calibration]\label{defn:calibration}
    A loss function $\ell'(f; \cD)$ is calibrated if its expected value $\EE_{\cD}[\ell'(f; \cD)]$ is minimized by the ground-truth CTR function $p^*$.
\end{definition}
 Based off of Definition~\ref{defn:calibration}, we first define a slightly relaxed notion of calibration, $\epsilon$-approximate calibration.
\begin{definition}[$\epsilon$-Approximate Calibration]\label{defn:approx_calibration}
    A loss function $\ell'(f; \cD)$ is said to be $\epsilon$-approximately calibrated if the expected value of the loss achieved by the ground-truth CTR function $p^*$ is at most $\epsilon$ greater than the minimum, namely
    \[
        \EE_{\cD}[\ell'(p^*; \cD)] - \min_{f \in\cH}\EE_{\cD}[\ell'(f; \cD)] \leq \epsilon.
    \]
\end{definition}
We then have the following guarantee for $\hat{\ell}_\sigma^{\textrm{hinge}, +}$.
\begin{theorem}\label{thm:plug_in_hinge_calibration}
    Let $\hat{p}$ be an estimate of $p^*$ such that $\EE[(\hat{p}(x) - p^*(x))^2] \leq \epsilon$. Assuming all bids are upper bounded by some $B \in \RR_{>0}$, for any $f \in \cH$ we have
    \begin{align*}
        &\EE_{\cD}[\mathrm{Welfare}_*(\cD)] \\
        &\qquad\leq \EE_{\cD}[\mathrm{Welfare}_f(\cD)] + \EE_{\cD}[\hat{\ell}^{\mathrm{hinge}, +}_\sigma(f; \cD)]\\
        &\hspace{3em} + \frac{n(n - 1)}{2}B\max\{1, \sigma B - 1\}\\
        &\hspace{3em} + n(n - 1)\sigma B^2 \sqrt{\epsilon} .
    \end{align*}
    Moreover, the loss function $\hat{\ell}_{\sigma}^{\textrm{hinge}, +}(f; \cD)$ is $\cO(\sqrt{\epsilon})$-approximately calibrated.
\end{theorem}
See Appendix~\ref{sec:proof_thm_plug_in_hinge_calibration} for detailed proof.
An important feature Theorem~\ref{thm:plug_in_estimate} and Theorem~\ref{thm:plug_in_hinge_calibration} share is that they depend only on the $\ell_2$ generalization error of the teacher network, and not on the explicit parametric assumptions on the distribution of eCPM. In other words, for any $\hat{p}$ we can simply use off-the-shelf results on its generalization error to show that using the induced $\hat{\ell}_{\sigma}^{\textrm{hinge}, +}(f; \cD)$ is approximately calibrated, with only the mild assumption that the CPC bids are bounded.

\subsection{Weighing with Teacher Networks}
\label{sec:weighted_variants}
The inclusion of the teacher network further guides us in developing theoretically-inspired weights for the proposed losses.
The goal of the weight for the pair $i, j$ is to mimic the indicator product $\ind\{i = i^*\}\ind\{j = j^*\}$, where $i^* = \argmax_{i \in [n]} b_ip_i$ and $j^* = \argmax_{i \in [n]} b_jf(x_j)$, so that the resulting loss better resembles the welfare suboptimality in~\eqref{eq:pair_welfare_loss}. The first indicator corresponds to the event that the ad $i$ has the highest ground-truth eCPM and the second the event that the ad $j$ has the highest predicted eCPM.
The weight should then be increasing in both $b_i\hat{p}(x_i)$ and $b_jf(x_j)$, with $\hat{p}$ being the teacher network.

\section{Experiments}
\label{sec:experiments}
We now demonstrate the advantages of our proposed losses on both simulated data and the Criteo Display Advertising Challenge dataset, a popular real-world benchmark for CTR prediction in ad auctions. Recalling Proposition~\ref{prop:calibrated_combo_loss}, we use the weighted sum of the logistic loss and the proposed ranking losses for all experiments to ensure the learned CTR model is sufficiently close to the ground truth.

\subsection{Synthetic Dataset}
For the simulation setting, we assume that the ads' features are 50-dimensional random vectors where each component is i.i.d. drawn from the standard normal distribution, namely $x_i \sim \cN(0, I_{50}),$ where $I_{50}$ denotes the 50-dimensional identity matrix.
For training, we generate 10,000 $x_i$'s from the $\cN(0, I_{50})$ distribution and generate the corresponding ground-truth CTR from a logistic model and the CPC bids from a log-normal distribution. We then draw the click indicators $y_i \sim \textrm{Ber}(p_i)$. We defer a more detailed description of the data-generating process to Appendix~\ref{sec:detailed_sim_setup}.

A two-layer neural network is used, where the hidden layer has 50 nodes with ReLU activation, and the output layer has one node with sigmoid activation. For evaluation, we simulate 2,000 auctions with 50 ads each. The training and evaluation processes are then repeated 30 times.

We begin by introducing the baselines we consider: logistic loss (also referred to as cross-entropy) and two versions of weighted logistic loss. Logistic loss is commonly used for training models for predicting CTRs, and is used by~\citet{guo2017deepfm,lian2018xdeepfm,chen2016deep} among many other works. Existing works~\citep{vasile2017cost,hummel2017loss} suggest the usage of a weighed logistic loss, with each entry weighted by the CPC bid. Finally, ~\citet{vasile2017cost} propose weighing the logistic loss by the square root of the CPC bid.

We focus on three loss function representative of what we proposed: $\ell^{\log}_{\sigma = 1}$, $\hat{\ell}_{\sigma = 1}^{\log}$, and $\hat{\ell}_{\sigma = 1}^{\textrm{hinge}, +}$. The first and the third correspond to~\eqref{eq:defn_log_surrogate} and~\eqref{eq:defn_hinge_surrogate_plug_in}, respectively. The second, $\hat{\ell}_{\sigma = 1}^{\log}$, replaces the $y_i$'s in $\ell^{\log}_{\sigma = 1}$ with $\hat{p}_i$ obtained from a teacher network. 

As discussed immediately after Theorem~\ref{thm:lower_bounds_welfare}, we add binary cross entropy loss to $\ell^{\log}_{\sigma = 1}, \hat{\ell}_{\sigma = 1}^{\log},$ and $\hat{\ell}_{\sigma = 1}^{\textrm{hinge}, +}$ and optimize over the composite loss. Additionally, motivated by Section~\ref{sec:weighted_variants}, we use logistic functions to weigh each pair in $\hat{\ell}_{\sigma = 1}^{\textrm{hinge}, +}$ and $\hat{\ell}^{\log}_{\sigma = 1}$. Both $\hat{\ell}_{\sigma= 1}^{\textrm{hinge}, +}$ and $\hat{\ell}_{\sigma = 1}^{\log}$ use the model trained with logistic loss as the teacher network. We defer a more detailed discussion to Appendix~\ref{sec:detailed_sim_setup}. 

\begin{figure}[th]
    \centering
    \includegraphics[width=\linewidth]{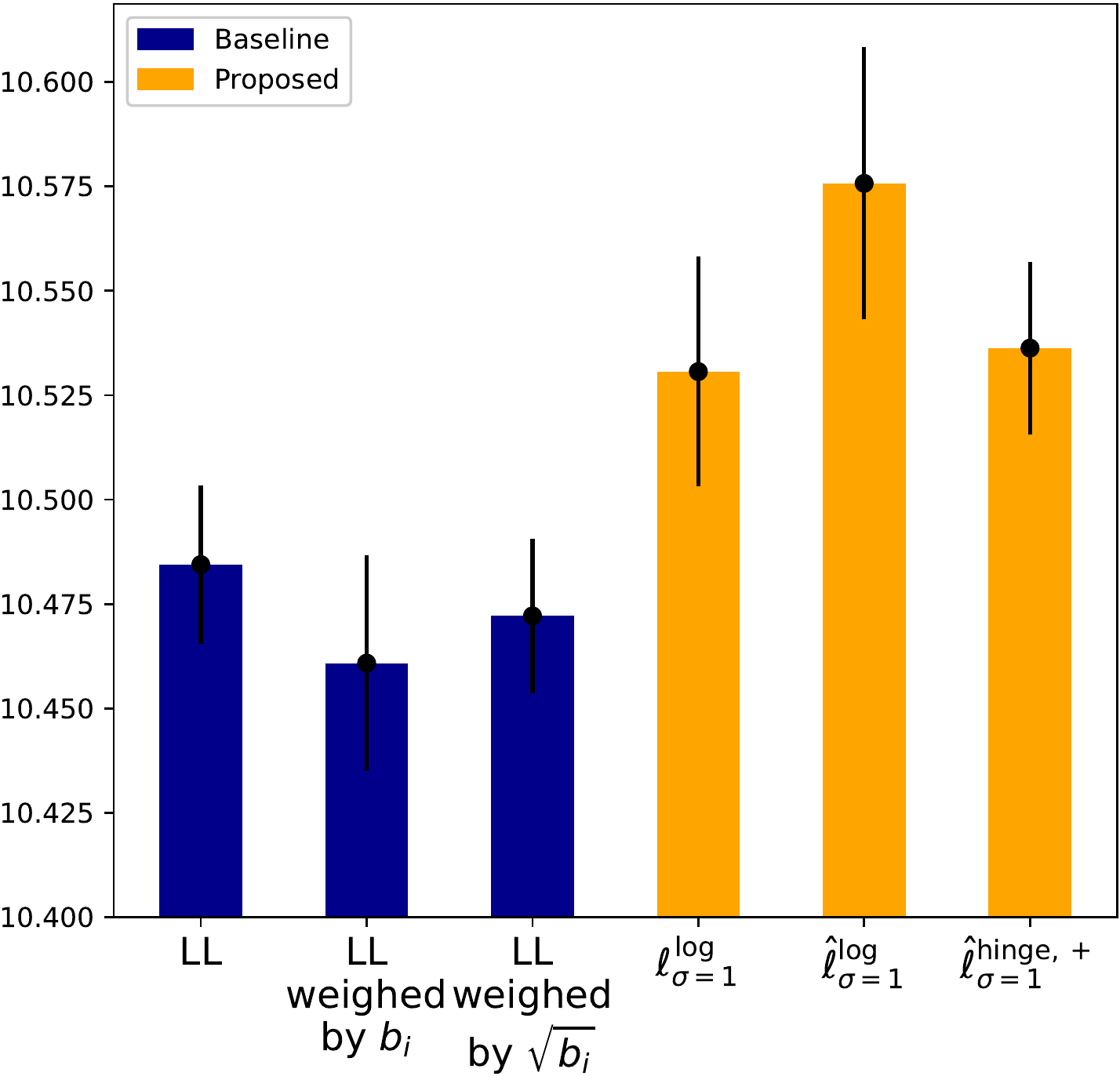}
    \vspace{-0.8cm}
    \caption{Test welfare on simulated data (higher is better). From left to right: (\textbf{In Blue}) models trained with logistic loss; logistic loss weighted by $b_i$~\citep{hummel2017loss}; logistic loss weighted by $\sqrt{b_i}$~\citep{vasile2017cost},  (\textbf{In Yellow}) proposed $\ell^{\log}_{\sigma = 1}$ indicator replaced by {logistic} function (defined in~\eqref{eq:defn_log_surrogate}); $\hat{\ell}^{\log}_{\sigma = 1}$ indicator replaced by {logistic} function + {student-teacher learning}; $\hat{\ell}_{\sigma = 1}^{\textrm{hinge}, +}$ indicator replaced by {hinge} function + {student-teacher learning} (defined in \eqref{eq:defn_hinge_surrogate_plug_in}).}
    \label{fig:sim_welfare}
\end{figure}

As we can see from Figure~\ref{fig:sim_welfare}, all three proposed pairwise ranking losses achieve higher test time welfare than the naive baselines. As we use the same model structure and optimizer for all models, it is further possible that with more careful tuning, the advantages of the pairwise ranking losses may be more pronounced.

{\noindent \textbf{Student-Teacher Learning.}} Comparing the performance of $\ell^{ \log}_{\sigma = 1}$ and $\hat{\ell}^{\log}_{\sigma = 1}$ shows that student-teacher learning overall beneficial for simulated data. Moreover, while $\hat{\ell}^{\rm hinge, +}_{\sigma = 1}$ is theoretically proven to be calibrated by Theorem~\ref{thm:plug_in_hinge_calibration}, in the simulated task we found that the loss does perform well compared to other proposed methods. We conjecture that the relatively modest performance is due to the fact that hinge function is not as smooth as logistic function, and thus is not well-suited for training neural networks.

{\noindent \textbf{Comparison with Existing Works.}} The experiment results also show that the loss functions derived in earlier works may depend on unrealistic assumptions and may be lacking in empirical justification, as can be seen in the performance of both weighted logistic losses. Regardless, we have shown that our proposed methods significantly outperform existing baselines.

\subsection{Criteo Dataset}

We use the popular Criteo Display Advertising Challenge dataset. We follow standard data preprocessing procedures and use a standard 8-1-1 train-validation-test split commonly found in the literature. We defer to Appendix~\ref{sec:detailed_criteo_setup} for a more detailed description of the setup.

We note there are several limitations to the dataset. Firstly, the Criteo dataset only includes ads that are shown. In an ad auction setting, this means that all ads have won their respective multi-slot auction. Moreover, the Criteo dataset only includes anonymous features, which means we have no access to key attributes such as the CPC bid or the slot for each ad. Lastly, we do not know the specific auction each ad belongs to. Unfortunately, these limitations are shared by all openly available benchmarks to the best of our knowledge.

For the first limitation, we note that it is near-impossible to learn accurate CTR models without assuming the CTRs of the shown ads follow the same distribution as those of the unshown ads. To handle the intrinsic bias between shown ads and unshown ads is very challenging and out of the scope of this paper. While the slot each ad belongs to is unavailable, as we argued previously, learning a position-based CTR model is not the focus of this work, and here we learn the CTR of each ad, assuming that it is assigned to the first slot. Finally, while we do not know the exact auction round, from Proposition~\ref{prop:welfare_maximization_reduction}, we know maximizing the welfare of multi-slot ad auctions requires maximizing the welfare of single-slot auctions over subsets of participating ads (given the position multipliers are independent wrt. advertisers). Thus, it remains viable to treat each minibatch as a specific instance of single-slot auction.

We generate the CPC bids using the outputs from a DeepFM model~\citep{guo2017deepfm} with randomly initialized weights, ensuring that the generated bids follow a log-normal distribution. Particularly, let $h(\cdot)$ denote a randomly initialized DeepFM model, we set $b_i = \exp(c \cdot h(x_i) + \epsilon_i)$, with $c$ being a scaling factor and $\epsilon_i$ a Gaussian noise.

We experiment with both DeepFM~\citep{guo2017deepfm} and DCN~\citep{wang2017deep}, two popular models with great performance on the Criteo dataset whose parameter choices are well-documented. To ensure that our loss functions benefit welfare \emph{holding all else constant}, we did not perform any parameter tuning or architecture search and used the model architectures and training protocols specified in the papers.

In this setting, we omit  $\hat{\ell}^{\textrm{hinge}, +}_\sigma$ as it is non-smooth and not well suited for complex neural network architectures considered here, given our empirical study.
Logistic loss is chosen as the baseline and $\ell_{\sigma = 3}^{\log}$ and $\hat{\ell}_{\sigma = 3}^{\log}$ are the proposed candidates due to smoothness. Here we set $\sigma = 3$ to better mimic the shape of the indicator variable. We omit weighted logistic losses proposed in~\citet{vasile2017cost,hummel2017loss} due to their poor performance on the synthetic dataset. We compare the losses based on three metrics: test-time welfare, area-under-curve (AUC) loss, and logistic loss, where AUC loss is defined as $1- \textsc{AUC}$.

For both DeepFM and DCN, we repeat the following procedure 10 times. We fit the models using logistic loss and $\ell_{\sigma = 3}^{\log}$. The model fitted using logistic loss is then used as the teacher network, whose outputs are used to construct $\hat{\ell}_{\sigma = 3}^{\log}$. We then evaluate the welfare, AUC loss, and logistic loss of the three models on the test set.

\begin{table}[htb]
    \centering
    \small
    \begin{tabular}{*3c}
        \toprule
        \multicolumn{3}{c}{DeepFM}\\
        \midrule
        {}   & Welfare   & AUC Loss \\
        LL (baseline)   &  $1.4448\pm 0.0025$ & $0.2200\pm 0.0004$  \\
        $\ell^{\log}_{\sigma = 3}$   &  $1.4622\pm 0.0021$ & $0.2169\pm 0.0003$ \\
        $\hat{\ell}^{\log}_{\sigma = 3}$   &  $\mathbf{1.4660\pm 0.0022}$  &  $0.2229\pm 0.0004$ \\
        \bottomrule
    \end{tabular}
    \caption{Welfare (higher is better) and AUC loss (lower is better) for DeepFM. Top to bottom: (\textbf{Baseline}) logistic loss; (\textbf{Proposed}) $\ell^{\log}_{\sigma = 3}$ indicator replaced by {logistic} function (defined in~\eqref{eq:defn_log_surrogate}); $\hat{\ell}^{\log}_{\sigma}$ indicator replaced by {logistic} function + {student-teacher learning}.}
    \label{tab:deepfm_short}
\end{table}

We report the welfare and AUC loss for DeepFM in Table~\ref{tab:deepfm_short}and those for DCN in Figure~\ref{fig:dcn}. Additional results including comparisons on the wall-clock run time can be found in  Appendix~\ref{sec:detailed_criteo_setup}.
\begin{figure}[htb]
    \centering
    \includegraphics[width=0.45\linewidth]{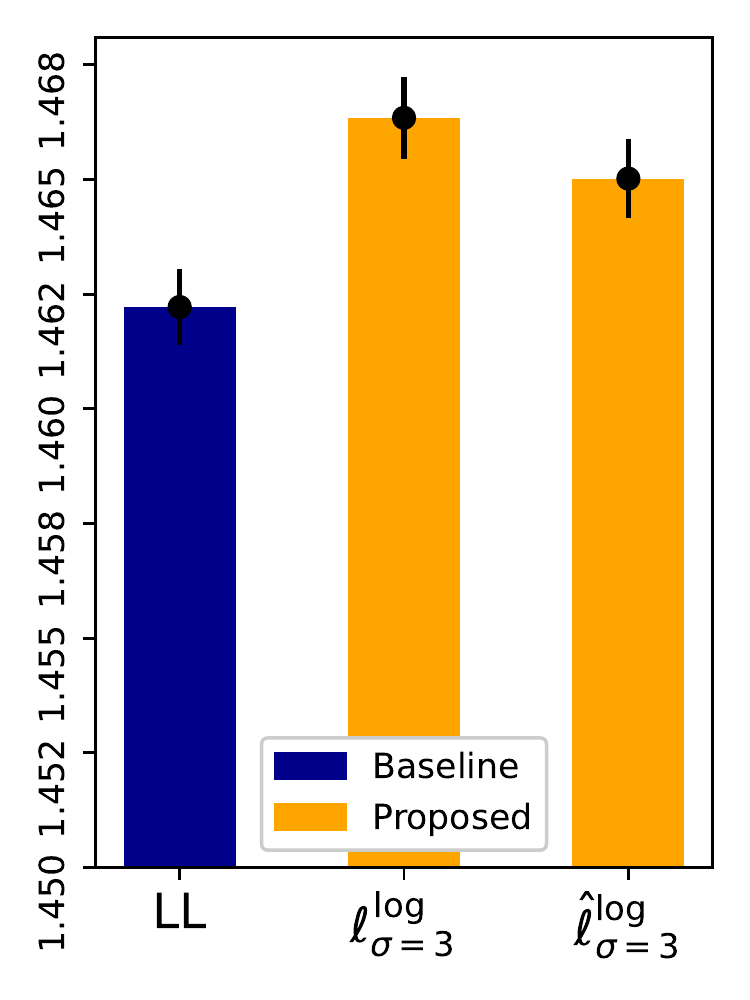}
    \hspace{0.03\linewidth}
    \includegraphics[width=0.45\linewidth]{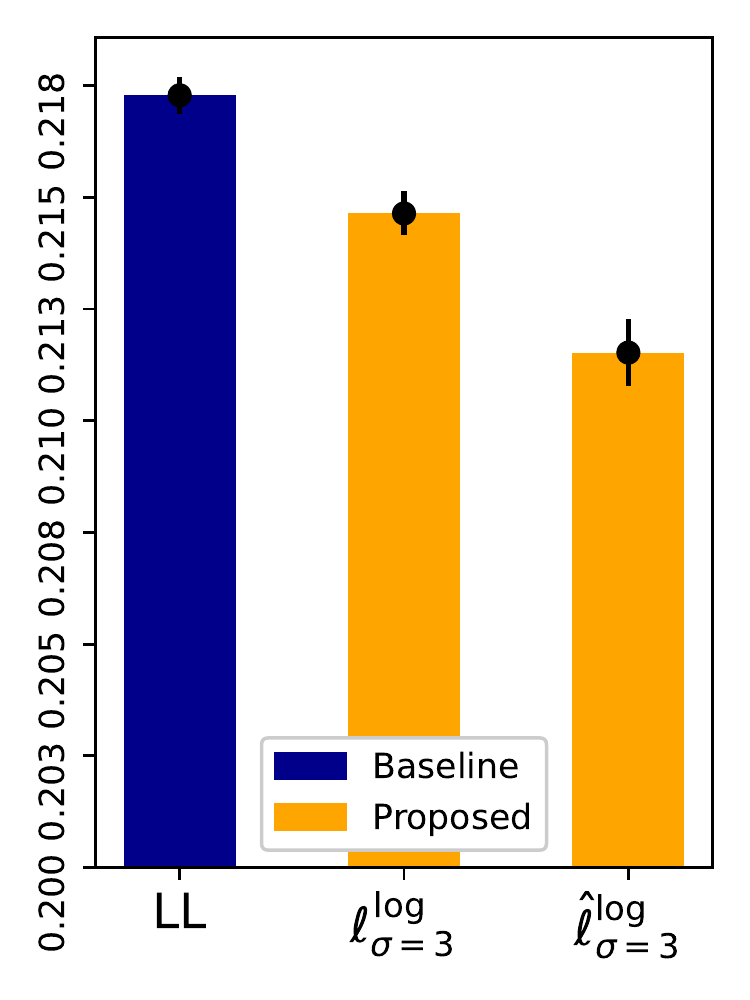}
    \caption{Experimental results for DCN. {\bf Left: Welfare} (higher is better). {\bf Right: AUC loss} (lower is better). For each plot, from left to right: (\textbf{Baseline, Blue}) logistic loss, (\textbf{Proposed, Yellow}) $\ell^{\log}_{\sigma = 3}$ indicator replaced by \emph{logistic} function (defined in~\eqref{eq:defn_log_surrogate}); $\hat{\ell}^{\log}_{\sigma}$ indicator replaced by \emph{logistic} function + \emph{student-teacher learning}.}
    \label{fig:dcn}
\end{figure}

As we observe from Table~\ref{tab:deepfm_short} and Figure~\ref{fig:dcn}, the proposed losses significantly improve test time welfare at a minimal cost (if any) to classification performance. Moreover, the improvement does not depend on the specific underlying model structure, and student-teacher learning continues to prove to be beneficial. Surprisingly, the proposed losses may also benefit AUC, a classification metric. We conjecture the improvement is due to the ranking loss formulation, which forces the model to better differentiate the ads' CTRs.

\section{Conclusion and Future Work} We propose surrogates that improve welfare for ad auctions with theoretical guarantees and good empirical performance. We hypothesize that the improvements will be more pronounced if we further tune the model architecture for the proposed losses and we leave architecture search as a future direction.


\section*{Acknowledgements}
Part of the work was completed while Boxiang Lyu and Zachary Robertson were Student Researchers at Google Research Mountain View. We would like to thank for Phil Long for the initial discussions, Ashwinkumar Badanidiyuru, Zhuoshu Li, and Aranyak Mehta for the insightful feedback.
\bibliographystyle{plainnat}
\bibliography{bib}
\appendix
\onecolumn
\section{Omitted Proofs}
\label{sec:omitted_proofs}
\subsection{Proof of Proposition~\ref{prop:welfare_maximization_reduction}}
\label{sec:proof_prop_welfare_maximization_reduction}
In order to prove Proposition~\ref{prop:welfare_maximization_reduction}, we begin with the following lemma that reduces welfare maximization to ranking.
\begin{lemma}\label{lemma:welfare_maximization_ranking}
    The function $f$ maximizes welfare in a $K$-slot auction only if it $\pi_f(k) = \pi_*(k)$ for all $k = 1, \ldots, K$.
\end{lemma}
\begin{proof}[Proof of Lemma~\ref{lemma:welfare_maximization_ranking}]
    We begin by showing a basic fact that ads with higher ground-truth eCPM should be ranked higher. Let $1 \leq k_1 \leq  k_2 \leq K$ be two arbitrary and fixed slots and $i, j$ be a pair of arbitrary and fixed ads, and assume without loss of generality that $b_ip_i \geq b_jp_j$. Immediately, we know that
    \[
        \alpha_{k_1} - \alpha_{k_2} \geq 0 \qquad  b_ip_i - b_jp_j \geq 0
    \]
    where the first inequality comes from the assumption that $\alpha_k$ is decreasing in $k$ and $k_1 \leq k_2$. The product of two nonnegative reals is nonnegative, and therefore
    \[ 
        (\alpha_{k_1} - \alpha_{k_2})(b_ip_i - b_jp_j) \geq 0
    \]
    which rearranges to
    \begin{equation*}
        \alpha_{k_1}b_ip_i + \alpha_{k_2}b_jp_j \geq \alpha_{k_1}b_jp_j + \alpha_{k_2}b_ip_i.
    \end{equation*}
    In other words, for pair of ads $(i, j)$ in the top $k$ slots, if $b_ip_i \geq b_jp_j$, then ad $i$ should be assigned to a slot that is closer to the first. As $p_i = p^*(x_i)$, the function $p^*$ correctly ranks each ad and therefore achieves maximum welfare.

    We now show $f$ maximizes welfare only if $\pi_f(k) = \pi_*(k)$ for all $k$.
    Let some CTR prediction function $f$ be arbitrary and fixed and assume there exists some $k_0$ where $\pi_f(k_0) \neq \pi_*(k_0)$. We can then divide the problem into two cases.
    \begin{enumerate}
        \item When $b_{\pi_f(k_0)}p_{\pi_f(k_0)}$ does not have top $K$ ground-truth eCPM. In other words, there is no $1 \leq k \leq K$ such that
        \[
            \pi_f(k_0) = \pi_*(k).
        \]
        In this case, the function $f$ has erroneously selected an ad who does not have top $K$ ground-truth eCPM and assigned it to the $k_0$-th slot. Moreover, the inclusion of the ad in the $K$-slots must imply that one of the ads with top $K$ ground-truth eCPM must be omitted by $f$. It is easy to see that replacing ad $\pi_f(k_0)$ with the ad that is erroenously left out improves $f$'s welfare.
        \item When $b_{\pi_f(k_0)}p_{\pi_f(k_0)}$ has top $K$ ground-truth eCPM. In other words, there is some $1 \leq k_1 \leq K$ such that $\pi_f(k_0) = \pi_*(k_1)$ where $k_1 \neq k_0$. The case can be further divided into two subcases.
        \begin{itemize}
            \item When $k_0 < k_1$. In this case, the ad is ranked higher by $f$ than it actually is. Similar to our reasoning for case 1, there must be an ad with top $k_0$ ground-truth eCPM that is erroneously left out of top $k_0$ by $f$. Switching ad $\pi_f(k_0)$ and the ad that is left out increases welfare.
            \item When $k_0 > k_1$. In this case, the ad is ranked lower by $f$ than it actually is. Therefore, there must be an ad with lower ground-truth eCPM in front of ad $\pi_f(k_0)$, and switching the two ads also increase welfare.
        \end{itemize}
    \end{enumerate}
    As we can see, whenever $\pi_f$ and $\pi_*$, there is always a method to improve the welfare achieved by $f$. Hence, $f$ maximizes welfare only if $\pi_f(k) = \pi_*(k)$ for $1 \leq k \leq K$.
\end{proof}

We then proceed with the proof of the proposition itself.
\begin{proof}[Proof of Proposition~\ref{prop:welfare_maximization_reduction}]
    Let $\{(b_i, x_i)\}_{i = 1}^n$ be the ads participating in the $K$-slot auction. Let $f$ denote an arbitrary and fixed CTR function. For all $1 \leq k \leq K - 1$, we define the set
    \[
        S_k = \{(b_i, x_i): b_ip^*(x_i) < b_{\pi_*(k)}p_{\pi_*(k)}\}.
    \]
    In other words, $S_k$ is the set of ads whose ground-truth eCPMs are \emph{outside} of top-$k$, excluding the ad with the $k$-th highest ground-truth eCPM. With a slight abuse of notation let 
    \[
        S_0 = \{(b_i, x_i)\}_{i = 1}^n.
    \]
    Observe that $f$ maximizes the welfare of the \emph{single}-slot auction over $S_k$ if and only if
    \[
        \pi_f(k + 1) = \pi_*(k + 1)
    \]
    for all $k = 1, \ldots, K - 1$, as we recall the construction of $S_k$ and Lemma~\ref{lemma:welfare_maximization_ranking}. Additionally, note that maximizing the welfare of the single-slot ad auction over $S_0$ requires $\pi_f(1) = \pi_*(1)$. Combine both observations, we know that $f$ maximizes the welfare of single-slot ad auctions over $S_0, \ldots, S_K$ if and only if $\pi_f(k) = \pi_*(k)$ for $k = 1, \ldots, K$. 

    Additionally, we know by Lemma~\ref{lemma:welfare_maximization_ranking} that $f$ maximizes the welfare of the $K$-slot auction over $S_0$ if and only if 
    \[
        \pi_f(k) = \pi_*(k), \quad \forall \, 1 \leq k \leq K.
    \]

    The two claims are thus equivalent and we complete the proof.
\end{proof}

\subsection{Proof of Proposition~\ref{thm:recovers_correct_ranking}}
\label{sec:proof_thm_recovers_correct_ranking}
\begin{proof}
    Let $\{(b_{i}, x_{i})\}_{i = 1}^{n}$ be arbitrary and fixed.
    Consider some pair \((i, j)\) and an arbitrary, not necessarily optimal, pCTR function $f$. We write out the parts of the conditional risk that depend only on the pair
    \begin{equation}
    \begin{split}
        &(b_{i}p_{i} - b_{j}p_{j})(\ind\{b_{i}f(x_{i}) \leq b_{j}f(x_{j})\} - (1 - \ind\{b_{i}f(x_{i}) \leq b_{j}f(x_{j})\}))\\
        &\qquad = (b_{i}p_{i} - b_{j}p_{j})(2\ind\{b_{i}f(x_{i}) \leq b_{j}f(x_{j})\} - 1)
    \end{split}
    \end{equation}
    The conditional risk for the pair is then minimized when $\ind\{b_{i}p_{i} \leq b_{j}p_{j}\} = \ind\{b_{i}f(x_{i}) \leq b_{j}f(x_{j})\}$. By the range of indicator variable, we know that
    \begin{equation*}
        (b_{i}p_{i} - b_{j}p_{j})(2\ind\{b_{i}f(x_{i}) \leq b_{j}f(x_{j})\} - 1) \geq -|b_{i}p_{i} - b_{j}p_{j}|.
    \end{equation*}
    Summing over all pairs $i, j$, the bound above implies
    \[
        \cR(f; \cD) \geq -\sum_{i \neq j}|b_{i}p_{i} - b_{j}p_{j}|.
    \]
    As we focus on the realizable setting, $p^* \in \cH$, and therefore
    \begin{align*}
        \min_{f \in \cH}\cR(f; \cD) \leq \cR(p^*) \leq -\sum_{i \neq j}|b_{i}p_{i} - b_{j}p_{j}|,
    \end{align*}
    which immediately implies that for any $\{(b_{i}, x_{i})\}_{i = 1}^{n}$,
    \[
        \min_{f \in \cH}\cR(f; \cD) = -\sum_{i \neq j}|b_{i}p_{i} - b_{j}p_{j}|.
    \]

    From realizability, there always exist at least one hypothesis in $\cH$ that minimize the conditional risk. Moreover, as we can see from the equation above, the ground-truth CTR function minimizes the conditional risk. Lastly, note that the ground-truth CTR ranks every pair correctly, which then implies the minimizer of the conditional risk must also rank each pair correctly.
\end{proof}

\subsection{Proof of Theorem~\ref{thm:lower_bounds_welfare}}
\label{sec:proof_thm_lower_bounds_welfare}
\begin{proof}
    Let $f\in\cH$ be arbitrary and fixed. 
    We proceed by first writing out the welfare loss in a pairwise fashion. For convenience, let $i^* = \argmax_{i' \in [n]} b_{i'}p_{i'}$ denote the index with the highest ground-truth eCPM. For an auction with $n$ participants, the optimal welfare is $\max_{i'\in [n]}b_{i'}p_{i'}$, which expands to
    \[
        \mathrm{Welfare}_*(\cD) = \sum_{i = 1}^n b_ip_i\ind\{i = i^*\}.
    \]
    Similarly, the welfare achieved by pCTR rule $f$ expands to
    \[
        \mathrm{Welfare}_f(\cD) = \sum_{j = 1}^n b_jp_j\ind\{j = j^*\},
    \]
    where we let $j^* = \argmax_{j' \in [n]} b_{j'}f(x_{j'}).$
    The difference between them is
    \[
        \mathrm{Welfare}_*(\cD) - \mathrm{Welfare}_f(\cD) = \sum_{i \in [n]}\sum_{j \in [n]}\ind\{i = i^*\}\ind\{j = j^*\}(b_ip_i - b_jp_j).
    \] 
    Since $\ind\{j = j^*\} = 1$ implies $b_jf(x_j) \geq b_if(x_i)$, the difference equals to
    \begin{equation*}
        \begin{split}
            &\mathrm{Welfare}_*(\cD) - \mathrm{Welfare}_f(\cD) = \sum_{(i, j) \in [n]^2}\ind\{i = i^*\}\ind\{j = j^*\}\ind\{b_jf(x_j) \geq b_if(x_i)\}(b_ip_i - b_jp_j).
        \end{split}
    \end{equation*}
    Consider an arbitrary and fixed unordered pair $(i, j)$ . As the summation contains both $(i, j)$ and $(j, i)$, we write out the parts of ground-truth welfare loss that depend on only the pair,
    \begin{equation}
        \label{eq:pairwise_welfare_suboptimality_one_pair}
        \begin{split}
            &\ind\{i = i^*\}\ind\{j = j^*\}\ind\{b_{i}f(x_{i}) \leq b_{j}f(x_{j})\}(b_{i}p_{i} - b_{j}p_{j})\\
            &\qquad + \ind\{j = i^*\}\ind\{i = j^*\}\ind\{b_{j}f(x_{j}) \leq b_{i}f(x_{i})\}(b_{j}p_{j} - b_{i}p_{i}).
        \end{split}
    \end{equation}
    When $b_{i}p_{i} \geq b_{j}p_{j}$, the sum is upper bounded by
    \begin{align*}
        &\ind\{i = i^*\}\ind\{j = j^*\}\ind\{b_{i}f(x_{i}) \leq b_{j}f(x_{j})\}(b_{i}p_{i} - b_{j}p_{j})  \leq \ind\{b_{i}f(x_{i}) \leq b_{j}f(x_{j})\}(b_{i}p_{i} - b_{j}p_{j}).
    \end{align*}
    When $b_{j}p_{j} \geq b_{i}p_{i}$, the sum is upper bounded by
    \begin{align*}
        &\ind\{j = i^*\}\ind\{i = j^*\}\ind\{b_{j}f(x_{j}) \leq b_{i}f(x_{i})\}(b_{j}p_{j} - b_{i}p_{i}) \leq \ind\{b_{j}f(x_{j}) \leq b_{i}f(x_{i})\}(b_{j}p_{j} - b_{i}p_{i}).
    \end{align*}
    Combining the two halves, we know that omitting ties, \eqref{eq:pairwise_welfare_suboptimality_one_pair} is upper bounded by
    \begin{align*}
        &\max\{\ind\{b_{i}f(x_{i}) \leq b_{j}f(x_{j})\}(b_{i}p_{i} - b_{j}p_{j}), \ind\{b_{j}f(x_{j}) \leq b_{i}f(x_{i})\}(b_{j}p_{j} - b_{i}p_{i})\}.
    \end{align*} 
    Ignoring ties, we may divide the problem to four cases: when $b_i f(x_i) > b_j f(x_j)$ and $b_i p_i > b_jp_j$, when $b_i f(x_i) < b_j f(x_j)$ and $b_i p_i > b_jp_j$, when $b_i f(x_i) > b_j f(x_j)$ and $b_i p_i < b_jp_j$, and finally when $b_i f(x_i) < b_j f(x_j)$ and $b_i p_i < b_jp_j$. We can show that in all four cases,
    \begin{align*}
        &\max\{\ind\{b_{i}f(x_{i}) \leq b_{j}f(x_{j})\}(b_{i}p_{i} - b_{j}p_{j}), \ind\{b_{j}f(x_{j}) \leq b_{i}f(x_{i})\}(b_{j}p_{j} - b_{i}p_{i})\}\\
        &\qquad = \frac{1}{2}(\ind\{b_{i}f(x_{i}) \leq b_{j}f(x_{j}) \}(b_{i}p_{i} - b_{j}p_{j}) + \ind\{b_{j}f(x_{j}) \leq b_{i}f(x_{i})\}(b_{j}p_{j} - b_{i}p_{i})+ |b_{i}p_{i} - b_{j}p_{j}|).
    \end{align*}

    Summing over all $(i, j) \in [n]^2$ and dividing by two shows
    \begin{align*}
        \mathrm{Welfare}_*(\cD) - \mathrm{Welfare}_f(\cD) \leq \frac{1}{2} \cR(f; \cD) + \frac{1}{4}\sum_{(i,j)\in[n]^2}|b_{i}p_{i} - b_{j}p_{j}|.
    \end{align*} Rearranging the terms gives us              
    \[
        \mathrm{Welfare}_f(\cD) \geq -\frac{1}{2}\cR(f; \cD) + \mathrm{Welfare}_*(\cD) - \frac{1}{4}\sum_{(i,j)\in[n]^2}|b_{i}p_{i} - b_{j}p_{j}|.
    \]
    Notice that the term $\mathrm{Welfare}_*(\cD) -\frac{1}{4}\sum_{(i,j)\in[n]}|b_{i}p_{i} - b_{j}p_{j}|$ is independent of $f$. Thus, we have shown that the conditional risk lower bounds $\mathrm{Welfare}_f$ up to some problem dependent constants and scaling.

    Recall from Appendix~\ref{sec:proof_thm_recovers_correct_ranking} $\min_{f\in\cH}\cR(f; \cD) = -\frac{1}{2}\sum_{i = 1}^n\sum_{j = 1}^n \abr{b_ip_i - b_jp_j}$. Let $\hat{f}$ be an arbitrary minimizer of $\cR(f; \cD)$ and we know
    \begin{align*}
        \mathrm{Welfare}_{\hat{f}}(\cD) \geq -\frac{1}{2} \cR(\hat{f}; \cD) + \mathrm{Welfare}_*(\cD) - \frac{1}{4}\sum_{i = 1}^n\sum_{j = 1}^n \abr{b_ip_i - b_jp_j}= \mathrm{Welfare}_*(\cD).
    \end{align*} 
    Recall that $\mathrm{Welfare}_*(\cD)$ is the maximum welfare achievable by definition. Thus the inequality is tight for any maximizer of $\cR(f; \cD)$.
\end{proof}

\subsection{Proof of Proposition~\ref{prop:calibrated_combo_loss}}
\label{sec:proof_prop_calibrated_combo_loss}
\begin{proof}
    We begin by showing $p^*$ is a minimizer of $\EE_{\cD}[\ell(f;\cD) + \lambda h(f; \cD)]$. By Proposition~\ref{thm:recovers_correct_ranking}, we know that $p^*$ minimizes $\cR(f; \cD)$ for all $\cD$ and therefore, $p^*$ minimizes $\EE[\ell(f; \cD)]$ by extension. The claim then easily hold.

    We then show that $p^*$ is unique. Let $f' \neq p^*$ be arbitrary and fixed. Because $p^*$ is the unique minimizer of $\EE_{\cD}[h(f; \cD)]$
    \[
        \EE_{\cD}[h(f'; \cD)] > \EE_{\cD}[h(p^*; \cD)],
    \]
    and we emphasize that the inequality is strict. Moreover, because $p^*$ is the minimizer of $\cR(f; \cD)$ for all $\cD$, 
    \[
        \EE_{\cD}[\ell(f'; \cD)] \geq \EE_{\cD}[\ell(p^*; \cD)].
    \]
    As $\lambda > 0$, we know $p^*$
    \[
        \EE_{\cD}[\ell(f'; \cD) + \lambda h(f'; \cD)] > \EE_{\cD}[\ell(p^*; \cD) + \lambda h(p^*; \cD)]
    \]
    and hence $p^*$ is the unique minimizer.
\end{proof}
\subsection{Proof of Theorem~\ref{thm:log_surrogate}}
\label{sec:proof_thm_smooth_surrogate}


We first show the claim holds for $\ell^{\log}_\sigma(f; \cD)$.
\begin{proof}[Proof for $\ell^{\log}_\sigma(f; \cD)$.]
    Consider an arbitrary pair $(i,j)$. For simplicity, let $\Delta_{ij} = b_ip_i - b_jp_j$ and $\Delta^f_{ij} = b_if(x_i) - b_jf(x_j)$. 
    We begin by writing out the parts of $\ell^{\log}_{\sigma}(f)$ and $\ell(f)$ that depend only on the pair $(i,j)$, which are
    \begin{equation}
    \label{eq:one_pair_log_logistic_loss}
        \Delta_{ij}\log(1 + \exp(-\sigma \Delta^f_{ij})) - \Delta_{ij}\log(1 + \exp(\sigma \Delta^f_{ij}))
    \end{equation} and 
    \begin{equation}
    \label{eq:one_pair_original_loss}
        -\Delta_{ij}\ind\{\Delta^f_{ij} \geq 0\} + \Delta_{ij}\ind\{\Delta^f_{ij} \leq 0\}, 
    \end{equation}
    respectively. Without loss of generality we assume throughout the rest of this proof that $\Delta^f_{ij} \geq 0$ and know that
    \begin{align*}
        \eqref{eq:one_pair_log_logistic_loss} - \eqref{eq:one_pair_original_loss} &= \Delta_{ij} \rbr{\log(1 + \exp(-\sigma \Delta^f_{ij})) - \log(1 + \exp(\sigma \Delta^f_{ij})) + 1},
    \end{align*}
    which in turn implies
    \begin{align*}
        |\eqref{eq:one_pair_log_logistic_loss} - \eqref{eq:one_pair_original_loss}|&\leq |\Delta_{ij}|\abr{\log(1 + \exp(-\sigma \Delta^f_{ij})) - \log(1 + \exp(\sigma \Delta^f_{ij})) + 1}.
    \end{align*}
    
    We now focus on the function $g(x) = \log(1 + \exp(-\sigma x)) - \log(1 + \exp(\sigma x)) + 1$. We quickly note that the function is monotonically decreasing in $x$ and hence
    \begin{align*}
        \sup_{x \in [0, B]}|g(x)| \leq \max\{1,  |\log(1 + \exp(-\sigma B)) - \log(1 + \exp(\sigma B)) + 1|\}
    \end{align*} Since $\sigma, B > 0$, we always have $\log(1 + \exp(\sigma B)) - \log(1 + \exp(-\sigma B)) \geq 0$. Divide the problem to two cases.
    \begin{enumerate}
        \item When $\log(1 + \exp(\sigma B)) - \log(1 + \exp(-\sigma B)) \in [0, 2)$. In this case we always have
            \begin{align*}
                &|\log(1 + \exp(-\sigma B)) - \log(1 + \exp(\sigma B)) + 1| \\
                &\qquad = \abr{1 - (\log(1 + \exp(\sigma B)) - \log(1 + \exp(-\sigma B)))}\\ &\qquad\leq 1
            \end{align*}
            and thus $|g(x)| \leq 1$ for all $x \in[0, B]$.
        \item When $\log(1 + \exp(\sigma B)) - \log(1 + \exp(-\sigma B)) \geq 2$, we have
            \[
                \sup_{x\in[0,B]}|g(x)| =  \log(1 + \exp(\sigma B)) - \log(1 + \exp(-\sigma B)) - 1.
            \]
    \end{enumerate}
    Combine the two cases above, and we have for all $x \in [0, B]$, $g(x) \leq \max\{1, \log(1 + \exp(\sigma B) - \log(1 + \exp(-\sigma B)) - 1\}$. Recall that all bids are upper bounded by $B$ and $f: \cX \to [0, 1]$, and we know $\Delta^f_{ij} \in [0, B]$ under our assumption. Therefore, for any arbitrary pair $(i,j)$ we always have
    \begin{align*}
        \eqref{eq:one_pair_log_logistic_loss} - \eqref{eq:one_pair_original_loss} & \leq\Delta_{ij} \rbr{\log(1 + \exp(-\sigma \Delta^f_{ij})) - \log(1 + \exp(\sigma \Delta^f_{ij})) + 1}\\
        &\leq |\Delta_{ij}|\max\{1, \log(1 + \exp(\sigma B)) - \log(1 + \exp(-\sigma B)) - 1\},
    \end{align*} and summing the equation over all $\frac{n(n - 1)}{2}$ unique pairs gives us
    \begin{align*}
        |\ell_{\sigma}^{\log}(f) - \ell(f)| \leq \frac{1}{2}\max\{1, \log(1 + \exp(\sigma B)) - \log(1 + \exp(-\sigma B)) - 1\}\sum_{i = 1}^n\sum_{j = 1}^n|b_ip_i - b_jp_j|.
    \end{align*}
    
    Finally, rearrange the term $\log(1 + \exp(\sigma B)) - \log(1 + \exp(-\sigma B))$ and we have the equation
    \[
        \log\rbr{\frac{1 + \exp(\sigma B)}{1 + \exp(-\sigma B)}} = 2,
    \]
    which solves to $\sigma B = 2$, namely $\sigma = 2/B$.
\end{proof}

The proof for $\ell^{\rm hinge}_\sigma (f; \cD)$ is similar.

\begin{proof}[Proof for $\ell^{\rm hinge}_\sigma(f; \cD)$.]
    Consider an arbitrary pair $(i,j)$. For simplicity, let $\Delta_{ij} = b_ip_i - b_jp_j$ and $\Delta^f_{ij} = b_if(x_i) - b_jf(x_j)$. 
    We begin by writing out the parts of $\ell^{\rm hinge}_{\sigma}(f)$ that depend only on the pair $(i,j)$, which are
    \begin{equation}
    \label{eq:one_pair_hinge_loss}
        \Delta_{ij}(-\sigma \Delta^f_{ij})_+ - \Delta_{ij}( \sigma \Delta^f_{ij})_+
    \end{equation}
    and recall from~\eqref{eq:one_pair_original_loss} that the corresponding part of $\ell(f)$ is
    \begin{equation*}
        -\Delta_{ij}\ind\{\Delta^f_{ij} \geq 0\} + \Delta_{ij}\ind\{\Delta^f_{ij} \leq 0\}.
    \end{equation*}
    Without loss of generality we assume that $\Delta^f_{ij} \geq 0$ and know that
    \begin{align*}
        \eqref{eq:one_pair_hinge_loss} - \eqref{eq:one_pair_original_loss} &= \Delta_{ij} \rbr{(-\sigma \Delta^f_{ij})_+ - ( \sigma \Delta^f_{ij})_+ + 1}= \Delta_{ij}\rbr{1-\sigma \Delta^f_{ij}}.
    \end{align*}
    Recall that $\Delta^f_{ij} \in [0, B]$. Therefore, for any arbitrary pair $(i,j)$ we always have
    \begin{align*}
        |\eqref{eq:one_pair_hinge_loss} - \eqref{eq:one_pair_original_loss}| \leq |\Delta_{ij}| \abr{1-\sigma \Delta^f_{ij}} \leq |\Delta_{ij}|\max\{1, 1 - \sigma B\}
    \end{align*} 
    and summing the equation over all $\frac{n(n - 1)}{2}$ unique pairs and applying triangle inequality gives us
    \begin{align*}
        |\ell_{\sigma}^{\log}(f) - \ell(f)| \leq \frac{1}{2}\max\{1, \sigma B - 1\}\sum_{i = 1}^n\sum_{j = 1}^n|b_ip_i - b_jp_j|.
    \end{align*}
    Setting $\sigma = 2/B$ completes the proof.
\end{proof}

\subsection{Proof of Proposition~\ref{prop:direct_learning_to_rank_fails}}
\label{sec:proof_prop_direct_learning_to_rank_fails}
\begin{proof}
    Without loss of generality we restrict our focus to a pair of ads $i, j$ where $b_ip_i \geq b_jp_j$. We then know that
    \begin{align*}
        \Pr(\ind\{b_iy_i \geq b_jy_j\} \neq \ind\{b_ip_i \geq b_jp_j\}) &= \Pr(b_iy_i \leq b_jy_j)\\
        &= \Pr(y_i = 0 \land y_j = 1) = (1 - p_i)p_j.
    \end{align*}
    Set $p_j = 1 - \frac{1}{2}\epsilon$ and $p_i = 1 - \frac{\epsilon}{p_j}$. We first verify that $p_i, p_j \in (0, 1)$.
    \begin{itemize}
        \item For $p_j$, because $\epsilon \in \rbr{0, \frac{1}{2}}$, $p_j \in \rbr{\frac{3}{4}, 1}$ and is valid.
        \item For $p_i$, because $p_j \in \rbr{\frac{3}{4}, 1}$, $\frac{\epsilon}{p_j} \in \rbr{\epsilon, \frac{4}{3}\epsilon} \subset \rbr{0, \frac{2}{3}}$, which in turn shows $p_i \in (0, 1)$ and is valid.
    \end{itemize}
    Moreover, plug in the choices of $p_i$, $p_j$ and we have
    \[
        \Pr(\ind\{b_iy_i \geq b_jy_j\} = \ind\{b_ip_i \geq b_jp_j\}) = 1 - (1 - p_i)p_j = 1 - (1 - \epsilon) = \epsilon, 
    \]
    which is exactly what we wanted.
    
    Lastly, we note that setting $b_i = \frac{2C}{p_i}, b_j = \frac{C}{p_j}$ ensures that $b_ip_i \geq b_jp_j$ for any $C \in \RR_{> 0}$.
\end{proof}

\subsection{Proof of Theorem~\ref{thm:plug_in_estimate}}
\label{sec:proof_thm_plug_in_estimate}
\begin{proof}
    Let $\cD$ be an arbitrary and fixed dataset. We drop the notation $\cD$ for the rest of the proof.
    Consider some pair \((i, j)\) and an arbitrary, not necessarily optimal, pCTR function $f$. We write out the parts of the conditional risk $\cR(f; \cD)$ that depend only on the pair,
    \[
        -(b_{i}p^*(x_{i})-b_{j}p^*(x_{j}))\ind\{b_{i}f(x_{i}) > b_{j}f(x_{j})\} - (b_{j}p^*(x_{j})-b_{i}p^*(x_{i}))\ind\{b_{j}f(x_{j}) > b_{i}f(x_{i})\}.
    \] 
    Similarly, the parts of $\hat{\ell}(f)$ that depend only on the pair $(i, j)$ is 
    \[
        -(b_{i}\hat{p}(x_{i})-b_{j}\hat{p}(x_{j}))\ind\{b_{i}f(x_{i}) > b_{j}f(x_{j})\} - (b_{j}\hat{p}(x_{j})-b_{i}\hat{p}(x_{i}))\ind\{b_{j}f(x_{j}) > b_{i}f(x_{i})\}.
    \] Omitting ties, we know one and exactly one of $\ind\{b_{i}f(x_{i}) > b_{j}f(x_{j})\}$ and $\ind\{b_{j}f(x_{j}) > b_{i}f(x_{i})\}$ is non-zero. Without loss of generality we assume $b_i f(x_i) > b_jf(x_j)$. The absolute value of the difference between the two pairs is then 
    \begin{align*}
        |b_{i}\hat{p}(x_{i})-b_{j}\hat{p}(x_{j}) - b_{i}{p}^*(x_{i})-b_{j}{p}^*(x_{j})|&\leq b_i|\hat{p}(x_i) - p^*(x_i)| + b_j |\hat{p}(x_j) - p^*(x_j)|\\
        &\leq B|\hat{p}(x_i) - p^*(x_i)| + B|\hat{p}(x_j) - p^*(x_j)|,
    \end{align*} where for the second inequality we use the assumption that all bids are upperbounded by $B$. Summing the inequality above over all $\frac{n(n - 1)}{2}$ pairs gives us
    \begin{align*}
        \EE_{\cD}[|\hat{\ell}(f) - \cR(f; \cD)|] &\leq n(n - 1)B\EE_{x_i}[|\hat{p}(x_i) - p^*(x_i)|] \leq n(n - 1)B\sqrt{\epsilon},
    \end{align*} where we use Jensen's inequality for the second inequality, completing the proof.
\end{proof}

\subsection{Proof of Theorem~\ref{thm:plug_in_hinge_calibration}}
\label{sec:proof_thm_plug_in_hinge_calibration}
We begin with a slight detour and first prove the following mathematical proposition.
\begin{proposition}\label{prop:hinge_property}
    For any pair of real numbers $a, b \in \RR$, we have the following
    \begin{equation*}
        |a_+ - b_+| + |(-a)_+ - (-b)_+| = |a - b|.
    \end{equation*}
\end{proposition}
\begin{proof}
    We divide the proposition into four cases.
    \begin{itemize}
        \item When $a \geq 0, b \geq 0$. In this case left-hand side evaluates to $| a  - b |$ and the equation holds.
        \item When $a < 0, b \geq 0$. In this case left-hand side evaluates to $| b | + | a | = b - a = | a - b |$ and the equation holds.
        \item When $a \geq 0, b < 0$. In this case left-hand side evaluates to $| a | + | b | = a - b = | a - b |$ and the equation holds.
        \item When $a < 0, b < 0$. In this case left-hand side evaluates to $| -a - (-b) | = | a - b |$ and the equation holds.
    \end{itemize}
\end{proof}

We also make use of the following helper function. Let
\begin{equation}
\label{eq:defn_ell_plus}
    \ell^+(f; \cD) = \sum_{i = 1}^n\sum_{j = 1}^n (b_ip_i - b_jp_j)_+ \ind\{b_if(x_i) \leq b_jf(x_j)\}.
\end{equation}
At a high level, $\ell^+(f; \cD)$ is the version of $\ell(f; \cD)$ when we know exactly what $p^*(\cdot)$ is. The loss function is, unfortunately, difficult to estimate (recall Proposition~\ref{prop:direct_learning_to_rank_fails}), but remains tightly related to welfare. We have the following proposition, which remains useful for the rest of the section.
\begin{proposition}\label{prop:ell_plus_welfare}
    We have the following inequality for all $f \in \cH$ and $\cD$
    \begin{align*}
        \mathrm{Welfare}_f(\cD) &\geq \mathrm{Welfare}_*(\cD) - \ell^+(f; \cD).
    \end{align*}
     Moreover, the inequality is tight for any minimizer of $\ell^+(f; \cD)$.
\end{proposition}
\begin{proof}
    The proof is nearly the same as that of Theorem~\ref{thm:lower_bounds_welfare}, which we provided in Appendix~\ref{sec:proof_thm_lower_bounds_welfare}. We detail the proof below for completeness.

    Condition on an arbitrary and unordered pair of indices $(i, j)$ and assume without loss of generality that $b_ip_i \geq b_jp_j$. The parts of $\ell^+(f; \cD)$ that depend only on the pair
    \begin{equation}
        \label{eq:ell_plus_one_pair}
        (b_ip_i - b_jp_j)_+\ind\{b_if(x_i) \leq b_jf(x_j)\} + (b_jp_j - b_ip_i)_+\ind\{b_jf(x_j) \leq b_if(x_i)\} = (b_ip_i - b_jp_j)\ind\{b_if(x_i) \leq b_jf(x_j)\}. 
    \end{equation}
    Recalling~\eqref{eq:pairwise_welfare_suboptimality_one_pair}, the welfare suboptimality induced by the pair is exactly
    \begin{equation}
        \label{eq:welfare_one_pair_conditioned}
        (b_{i}p_{i} - b_{j}p_{j})\ind\{i = i^*\}\ind\{j = j^*\}\ind\{b_{i}f(x_{i}) \leq b_{j}f(x_{j})\},
    \end{equation}
    where we note the assumption that $b_ip_i \geq b_jp_j$ implies $j \neq i^*$, where we recall $i^* = \argmax_{i' \in [n]} b_{i'}p_{i'}$. Immediately we note that
    \begin{align*}
        \eqref{eq:welfare_one_pair_conditioned} &\leq \eqref{eq:ell_plus_one_pair} \leq \eqref{eq:welfare_one_pair_conditioned} + (b_ip_i - b_jp_j)_+.
    \end{align*}
    Particularly $\eqref{eq:ell_plus_one_pair} = \eqref{eq:welfare_one_pair_conditioned}$, when $b_if(x_i) \geq b_jf(x_j)$. We then sum over all pairs of $(i, j)$ and know that
    \begin{equation}
        \mathrm{Welfare}_*(\cD) - \mathrm{Welfare}_f(\cD) \leq \ell^+(f; \cD).
    \end{equation}
    and the inequality is tight when $b_if(x_i) \geq b_jf(x_j)$ whenever $b_ip_i \geq b_jp_j$. Since $p^* \in \cH$, it is possible to exactly minimize $\ell^+(f; \cD)$ for all $\cD$, thus any minimizer of $\ell^+(f; \cD)$ must rank each pair correctly. Therefore, the inequality is tight for any minimizer of $\ell^+(f; \cD)$ for any $\cD$.
\end{proof}

We now proceed with the proof of Theorem~\ref{thm:plug_in_hinge_calibration} itself.

\begin{proof}[Proof of Theorem~\ref{thm:plug_in_hinge_calibration}]
    Consider an arbitrary pair $(i,j)$. For simplicity, let $\Delta_{ij} = b_ip_i - b_jp_j$, $\hat{\Delta}_{ij} = b_i\hat{p_i} - b_j \hat{p_j}$, and $\Delta^f_{ij} = b_if(x_i) - b_jf(x_j)$. 
    We begin by writing out the part of $\hat{\ell}^{\mathrm{hinge}, +}_{\sigma}(f; \cD)$ that depend only on the pair $(i,j)$
    \begin{equation}
    \label{eq:one_pair_estim_hinge_loss}
        (\hat{\Delta}_{ij})_+(-\sigma \Delta^f_{ij})_+ + (-\hat{\Delta}_{ij})_+(\sigma \Delta^f_{ij})_+.
    \end{equation}
    We also introduce the loss function
    \[
        {\ell}_{\sigma}^{\mathrm{hinge}, +}(f; \cD) = \sum_{i = 1}^n \sum_{j = 1}^n (\Delta_{ij})_+ (-\sigma \Delta^f_{ij})_+,
    \] 
    and the part that corresponds to the pair $(i, j)$ would be
    \begin{equation}
    \label{eq:one_pair_true_hinge_loss}
        ({\Delta}_{ij})_+(-\sigma \Delta^f_{ij})_+ + (-{\Delta}_{ij})_+(\sigma \Delta^f_{ij})_+.
    \end{equation}
    We then know
    \begin{align*}
        |\eqref{eq:one_pair_estim_hinge_loss} - \eqref{eq:one_pair_true_hinge_loss}| &= |(-\sigma \Delta^f_{ij})_+((\hat{\Delta}_{ij})_+ - (\Delta_{ij})_+) + (\sigma \Delta^f_{ij})_+((-\hat{\Delta}_{ij})_+ - (-\Delta_{ij})_+)|\\
        &\leq \max\{(-\sigma \Delta^f_{ij})_+, (\sigma \Delta^f_{ij})_+\}(|(\hat{\Delta}_{ij})_+ - (\Delta_{ij})_+| + |(-\hat{\Delta}_{ij})_+ - (-\Delta_{ij})_+|)\\
        &= \max\{(-\sigma \Delta^f_{ij})_+, (\sigma \Delta^f_{ij})_+\}|\hat{\Delta}_{ij} - \Delta_{ij}|\\
        &\leq \sigma B |\hat{\Delta}_{ij} - \Delta_{ij}|,
    \end{align*}
    where for the second equality we use Proposition~\ref{prop:hinge_property} and the last inequality the fact that the bids are bounded by $B$. As $\EE[(\hat{p}(x) - p^*(x))^2] \leq \epsilon$,
    \begin{align*}
        \EE[(\hat{\Delta}_{ij} - \Delta_{ij})^2] &\leq 2\EE[(b_ip_i - b_i\hat{p}(x_i))^2] + 2\EE[(b_j p_j - b_j\hat{p}(x_j))^2]\\
        &\leq 2B^2\EE[(\hat{p}(x) - p^*(x))^2] + 2B^2\EE[(\hat{p}(x) - p^*(x))^2] \leq 4B^2\epsilon
    \end{align*} 
    where for the first inequality we recall the fact that $(a + b)^2 \leq 2a^2 + 2b^2$. By Jensen's inequality, we know
    \[
        \EE[|\hat{\Delta}_{ij} - \Delta_{ij}|] \leq 2B\sqrt{\epsilon}.
    \]
    
    Summing over all unique pairs of $(i, j)$ and we know for any $f$
    \begin{equation}
        \label{eq:relationship_hat_ell_ell}
        \EE_{\cD}[|\hat{\ell}^{\mathrm{hinge}, +}_\sigma(f; \cD) - \ell^{\mathrm{hinge}, +}_\sigma(f; \cD)|] \leq n(n - 1) \sigma B^2\sqrt{\epsilon}.
    \end{equation}
    
    We then control $\EE[|\ell^{\mathrm{hinge}, +}(f) - \ell^+(f)|]$ similar to how we proved Theorem~\ref{thm:log_surrogate}. Without loss of generality we assume that $\Delta_{ij} \geq 0$ and have
    \begin{align*}
        \eqref{eq:one_pair_true_hinge_loss} - \eqref{eq:ell_plus_one_pair} &= \Delta_{ij}((-\sigma \Delta^f_{ij})_+ - 1)
    \end{align*}
    and therefore
    \begin{align*}
        |\eqref{eq:one_pair_true_hinge_loss} - \eqref{eq:ell_plus_one_pair}| &\leq |\Delta_{ij}||(-\sigma \Delta^f_{ij})_+ - 1| \leq |b_ip_i - b_jp_j|\max\{1, \sigma B - 1\}.
    \end{align*}
    Summing the inequality across all pairs gives us
    \begin{align*}
        |{\ell}^{\mathrm{hinge}, +}_\sigma(f; \cD) - \ell^+f(; \cD)| &\leq \frac{1}{2}\max\{1, \sigma B - 1\}\sum_{i = 1}^n \sum_{j = 1}^n |b_ip_i - b_jp_j|\\
        &\leq \frac{n(n - 1)}{2} \max\{1, \sigma B - 1\}B.
    \end{align*}
    Taking the expectation over $\cD$ and applying Jensen's inequality, we then know that
    \begin{align*}
        \EE_{\cD}[|\hat{\ell}^{\mathrm{hinge}, +}_\sigma(f; \cD) - \ell^{+}(f; \cD)|] \leq \frac{n(n - 1)B}{2}(2\sqrt{\epsilon} + \max\{1, \sigma B - 1\}).
    \end{align*}
    By Proposition~\ref{prop:ell_plus_welfare}, we then know for any $f \in \cH$
    \begin{align*}
        \EE_{\cD}[\mathrm{Welfare}_f(\cD)] \geq \EE_{\cD}[\mathrm{Welfare}_*(\cD)] - \EE_{\cD}[\hat{\ell}^{\mathrm{hinge}, +}_\sigma(f; \cD)] - n(n - 1)\sigma B^2 \sqrt{\epsilon} - \frac{n(n - 1)}{2}B\max\{1, \sigma B - 1\}.
    \end{align*}
    
    We now show that $p^*$'s suboptimality tends to 0 as $\epsilon$ goes to 0. Recall the definition of $\ell^{\mathrm{hinge}, +}_\sigma(f; \cD)$, we have
    \[
        \ell^{\mathrm{hinge}, +}_\sigma (p^*; \cD) = \sum_{i = 1}^n\sum_{j = 1}^n (\Delta_{ij})_+(-\sigma \Delta_{ij})_+ = 0.
    \]
    Plug the result back to~\eqref{eq:relationship_hat_ell_ell}, and we know
    \[
        \EE_{\cD}[\hat{\ell}^{\mathrm{hinge}, +}_\sigma(p^*; \cD)] \leq n(n - 1) \sigma B^2 \sqrt{\epsilon}.
    \]
    Since $\hat{\ell}^{\mathrm{hinge}, +}_\sigma(f; \cD) \geq 0$ for any $f \in \cH$ and $\cD$, we have
    \[
        \EE_{\cD}[\hat{\ell}^{\mathrm{hinge}, +}_\sigma(p^*; \cD) - \min_f\hat{\ell}^{\mathrm{hinge}, +}_\sigma(f; \cD)] \leq n(n - 1) \sigma B^2 \sqrt{\epsilon}.
    \]
    Since $\min_f\EE_{\cD}[\hat{\ell}^{\mathrm{hinge}, +}_\sigma(f; \cD)] \leq \EE_{\cD}[\min_f\hat{\ell}^{\mathrm{hinge}, +}_\sigma(f; \cD)]$, we know that
    \[
        \EE_{\cD}[\hat{\ell}^{\mathrm{hinge}, +}_\sigma(p^*; \cD)] - \min_f\EE_{\cD}[\hat{\ell}^{\mathrm{hinge}, +}_\sigma(f; \cD)] \leq n(n - 1) \sigma B^2 \sqrt{\epsilon}
    \]
    and the loss is $\cO(\sqrt{\epsilon})$-approximately calibrated.
\end{proof}

\section{Detailed Description for Experiments on Synthetic Dataset}
\label{sec:detailed_sim_setup}

Here we include a detailed discussion on how we conducted the experiments on synthetic data.

\paragraph{Data Generating Process.} To generate the ground-truth CTRs, we draw a weight vector $w_p \in \mathbb{R}^{50}$ where each entry is follows a $\textrm{Unif}(-\sqrt{10}, \sqrt{10})$ distribution. We use a logistic model and set $p_i = \frac{1}{1 + \exp(w_p^Tx_i) + \xi^{\rm CTR}_i}$, where $\xi^{\rm CTR}$ is a random noise following $\cN(0, 0.1^2)$.

Similarly, in order to generate the bids, we draw a weight vector $w_b \in \mathbb{R}^{50}$ where each entry is drawn from $\textrm{Unif}(-2, 2)$. Inspired by the empirical observations made in~\citet{vasile2017cost}, we assume the CPC bids, when conditioned on the feature $x_i$, follow a log-normal distribution and set $b_i = \exp(w_b^Tx_i + \xi^{\rm bid}_i)$ where $\xi^{\rm bid}_i \sim N(0, 0.1^2)$. The choice of parameters avoids overflowing due to the exponential term used for generating the bids.

\begin{figure}[ht]
    \centering
    \includegraphics[width=0.8\linewidth]{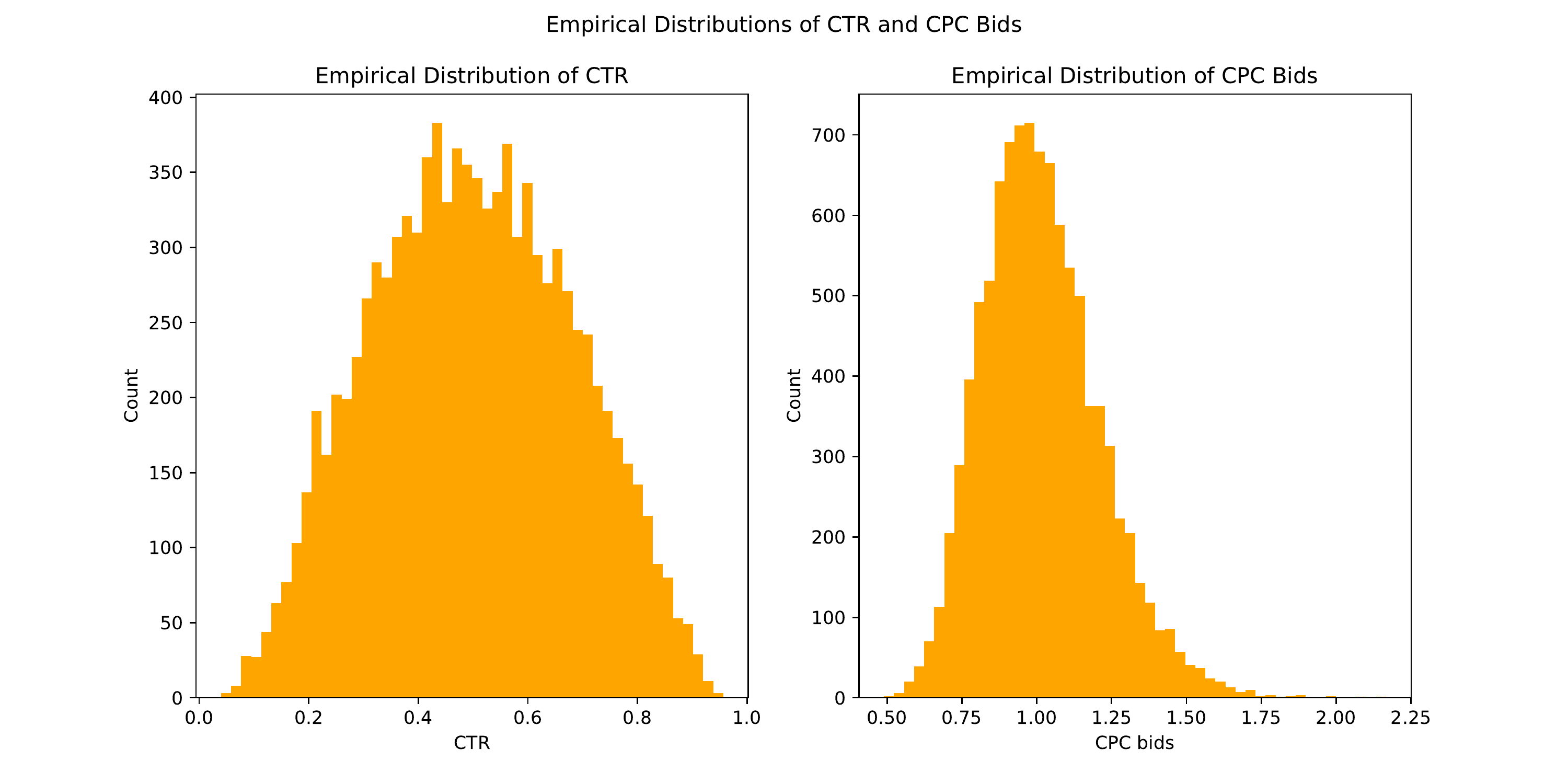}
    \caption{Distribution of the Generated CTR and CPC.}
    \label{fig:ctr_cpc_distribution}
\end{figure}

We visualize in Figure~\ref{fig:ctr_cpc_distribution} the distribution of the CTRs and CPC bids generated when $\sigma_{\rm CTR} = 0.1$ as a sanity check. We can verify that the generated CTRs is unimodal and centered around 0.5, and hence does not follow a degenerate distribution. The distribution of the generated CPC bids roughly follows a log-normal distribution, agreeing with the empirical observations made in~\citet{vasile2017cost}.

\paragraph{Training.} We used Adam with default parameters. Batch size is set to 256. All models are trained till convergence.

\paragraph{Baselines.} We use the logistic loss $\ell^{\rm LL}$ and weighted logistic loss $\ell^{\rm WLL}$ as baselines. Particularly, for any function $f$ and advertisement $((b_i, x_i), y_i)$, the logistic loss is
\[
    \ell^{\rm LL}(f(x_i), y_i) = -(y_i\log(f(x_i)) + (1 - y_i) \log(1 - f(x_i)))
\]
and the weighted logistic loss is
\[
    \ell^{\rm WLL}(f(x_i), y_i, b_i) = -b_i(y_i\log(f(x_i)) + (1 - y_i) \log(1 - f(x_i))).
\]
Summing them over all $((b_i, x_i), y_i) \in \cD$ yields the empirical loss.

\paragraph{Proposed Losses.} We begin by introducing the version of $\ell^{\log}_{\sigma}$ we used for the experiment setting, $\ell^{\log}_{\sigma = 1}$.
\[
    \ell^{\log}_{\sigma = 1}(f; \cD) =\sum_{i = 1}^n \sum_{j = 1}^n (b_iy_i - b_jy_j)\log(1 + \exp(-(b_if(x_i) - b_jf(x_j)))) + 3\sum_{i = 1}^n \ell^{\rm LL}(f(x_i), y_i).
\]
Intuitively, one can view $\ell^{\rm LL}(f(x_i), y_i)$ as a regularizer, avoiding the loss function to place too much emphasis on minimizing the ranking loss, and ensures our model can still adequately predict the CTRs. Similar to our choice of $\sigma = 1$, choosing 3 as the regularization strength of $\ell^{\rm}$ is arbitrary and we did not optimize over the space of possible regularization strengths. 

We then write out the version of $\hat{\ell}^{\textrm{hinge},+}_{\sigma}$ we employ, which, in addition to a logistic loss function as a regularizer, also uses logistic functions to weigh each pair. More formally, we have
\begin{align*}
    \hat{\ell}^{\textrm{hinge}, +}_{\sigma = 1}(f; \cD) = &\sum_{i = 1}^n \sum_{j = 1}^n \frac{1}{1 + \exp(-3b_i\hat{p}(x_i))}\frac{1}{1 + \exp(- 3b_jf(x_j))} (b_i\hat{p}_i - b_j \hat{p}_j)_+ (-(b_if(x_i) - b_jf(x_j)))_+\\
    &+ 3\sum_{i = 1}^n \ell^{\rm LL}(f(x_i), y_i).
\end{align*}

Particularly, here we use $\frac{1}{1 + \exp(-3b_i\hat{p}(x_i))}$ as a proxy to $\ind\{i = i^*\}$, which we recall indicates whether the $i$-th ad has the highest ground-truth eCPM or not. Similarly, $\frac{1}{1 + \exp(- 3b_jf(x_j))}$ is the proxy to $\ind\{j = j^*\}$. The choice to multiply $b_i\hat{p}(x_i)$ by 3 in the denominator is ad hoc and arbitrary. We did not tune over the space of possible indicators and leave a rigorous treatise of the parameter choice, under additional assumptions on the eCPMs' distribution, as an interesting future direction.

Finally, we write out $\hat{\ell}^{\log}_{\sigma = 1}(f; \cD),$ which uses the same weighing scheme used in $\hat{\ell}^{\textrm{hinge}, +}_{\sigma = 1}(f; \cD)$. We have
\begin{align*}
    \hat{\ell}^{\log}_{\sigma = 1}(f; \cD) = &\sum_{i = 1}^n \sum_{j = 1}^n \frac{1}{1 + \exp(-3b_i\hat{p}(x_i))}\frac{1}{1 + \exp(- 3b_jf(x_j))} (b_i\hat{p}(x_i) - b_j\hat{p}(x_j))\log(1 + \exp(-(b_if(x_i) - b_jf(x_j))))\\
    &+ 3\sum_{i = 1}^n \ell^{\rm LL}(f(x_i), y_i).
\end{align*}

For $\hat{\ell}^{\mathrm{hinge}, +}_{\sigma = 1}(f; \cD)$ and $\hat{\ell}^{\log}_{\sigma = 1}(f;\cD)$, the plug-in estimator $\hat{p}(\cdot)$ is model trained with logistic loss. We train on the baseline losses first, and then use the model outputs as the plug-in estimates for $\hat{\ell}^{\mathrm{hinge}, +}_{\sigma = 1}(f; \cD)$ and $\hat{\ell}^{\log}_{\sigma = 1}(f;\cD)$.

\paragraph{Evaluation.} As we know the exact CTR for each advertisement, we can directly calculate the social welfare achieved by the model trained with each of the five losses. For each model, we first pick out the ad with the highest predicted eCPM for each of the 2000 randomly generated auctions. We then calculate the actual CPM of the selected ad and record it as the social welfare achieved on that particular round of auction.

\paragraph{Calculating the Standard Error.} Since the bulk of the randomness in test time social welfare lies in the data generation process, reporting the standard error of each model's average test time social welfare would not be informative, as the data generating process' noise dominates. We instead report the standard error of \emph{the difference} between the particular model's social welfare and that of the average social welfare across the 5 models. The difference term is then able to account for the randomness within the data generating process, and we are measuring the standard error of the \emph{comparative} performance of each model.

\section{Detailed Description of Experiments on Criteo Dataset}
\label{sec:detailed_criteo_setup}
\paragraph{Data Preprocessing.} A standard practice is to preprocess the Criteo dataset using the approach used by the winners of the Criteo Display Advertising Challenge\footnote{\url{https://www.csie.ntu.edu.tw/~r01922136/kaggle-2014-criteo.pdf}}~\citep{chen2016deep,guo2017deepfm,lian2018xdeepfm,cheng2016wide}. The Criteo data contains 13 numerical feature columns with positive integer observations and we replace all observations with value greater than 2 with its log instead, namely
\[
    f(x) = \begin{cases}
        x &\textrm{ if $x \leq 2$}\\
        \lfloor \log_2(x) \rfloor&\textrm{ otherwise}
    \end{cases}.
\]
For the remaining 26 categorical features, we replace the values that appear less than 10 times with a special value. We also use a 8-1-1 train-validation-test split, commonly used by existing works.

\paragraph{Model Implementation.} To ensure our results can be easily reproduced, we use the open source implementation found in the DeepCTR package for both DeepFM and Deep \& Cross Network~\citep{shen2017deepctr}.

\paragraph{Bid Generation.} We use the DeepFM implementation found in~\citet{shen2017deepctr} with default parameters. Each categorical variable is cast to a 4 dimensional embedding. The weights are initialized randomly. The deep network has 3 hidden layers with decreasing size, consisting of 256, 128, and 64 nodes each. ReLU activation is used for all layers save for the output layer, which uses a sigmoid activation function. The processed data is then fed into the network and then re-scaled to the range $[0, 1]$ (otherwise the outputs are too close to zero). We then add a $\cN(0, 0.1^2)$ random noise and take the exponential over the scores. Letting $x$ denote the feature, we use
\[
    b_i = \exp(c \cdot f(x_i) + \xi^{\rm bid}_{i})
\]
as the CPC bids, where $c$ is a scaling constant ensuring $c f(x_i) \in [0, 1]$ and $\xi_{i}^{\rm bid} \sim \cN(0, 1)$.

\paragraph{Choice of Loss Functions.} We begin by introducing the version of $\ell^{\log}_{\sigma}$ used for the Criteo experiments, $\ell^{\log}_{\sigma = 3}$.
\begin{align*}
    \ell^{\log}_{\sigma = 3}(f; \cD) =&\sum_{i = 1}^n \sum_{j = 1}^n \frac{1}{1 + \exp(-3b_i\hat{p}(x_i))}\frac{1}{1 + \exp(- 3b_jf(x_j))}(b_iy_i - b_jy_j)\log(1 + \exp(-3(b_if(x_i) - b_jf(x_j)))) \\
    &+ \lambda  \sum_{i = 1}^n \ell^{\rm LL}(f(x_i), y_i).
\end{align*}
Compared to the version used for the simulation studies, for the Criteo simulations we further incorporate the weighing schemes discussed in Section~\ref{sec:weighted_variants}. The value of $\sigma$ is also adjusted slightly to better approximate the indicator function on the range of eCPMs in our setting.

The version of $\hat{\ell}^{\log}_{\sigma}$ used is defined as follows. We have
\begin{align*}
    \hat{\ell}^{\log}_{\sigma = 3}(f; \cD) = &\sum_{i = 1}^n \sum_{j = 1}^n \frac{1}{1 + \exp(-3b_i\hat{p}(x_i))}\frac{1}{1 + \exp(- 3b_jf(x_j))} (b_i\hat{p}(x_i) - b_j\hat{p}(x_j))\log(1 + \exp(-3(b_if(x_i) - b_jf(x_j))))\\
    &+ \lambda \sum_{i = 1}^n \ell^{\rm LL}(f(x_i), y_i),
\end{align*}
and we slightly adjusted the choice of $\sigma$.

We perform a grid search to adjust the value of $\lambda$ over the set $\{0.1, 1, 3, 5, 10\}$ based on the welfare, AUC loss, and logistic loss achieved by the different choices. We set $\lambda = 3$ for both DeepFM and DCN as the choice achieves significant increases in welfare with minor impact on AUC loss and logistic loss for both $\ell^{\log}_{\sigma = 3}$ and $\hat{\ell}^{\log}_{\sigma = 3}$.
\paragraph{Evaluation.} We take the test set and partition it into auctions with 256 bidders each. We use the models trained on the three losses to determine the winner of each auction, and record the \emph{realized eCPM} (the observed $b_iy_i$'s) as the welfare of the auction. As the ground-truth CTR of the ads are unknown, we are forced to use the realized eCPM as proxies for measuring welfare.

\paragraph{Parameter Settings for DeepFM.} We follow the optimal parameters specified in~\citet{guo2017deepfm} and construct a DeepFM with 3 hidden layers, each with 400 nodes using ReLU activation. Embedding dimension for the categorial variables is set to 10. Dropout rate is set to 0.5. For optimizing the model we used Adam with default parameters and set batch size to 256. The model is trained for 3 epochs.

\paragraph{Parameter Settings for DCN.} We use the parameters set in~\citet{wang2017deep} and construct a DCN with 2 hidden layers, each with 1024 nodes using ReLU activation. Batch normalization is applied to the network and the number of cross layers is set to 6. For the categorical features, we set the dimensionality of the embedding as $6 \times (\textrm{feature cardinality})^{1/4}$. For optimizing the model we used Adam and set the batch size to 512. The model is trained for 150,000 steps.

\begin{figure}[ht]
    \centering
    \includegraphics[width=0.25\linewidth]{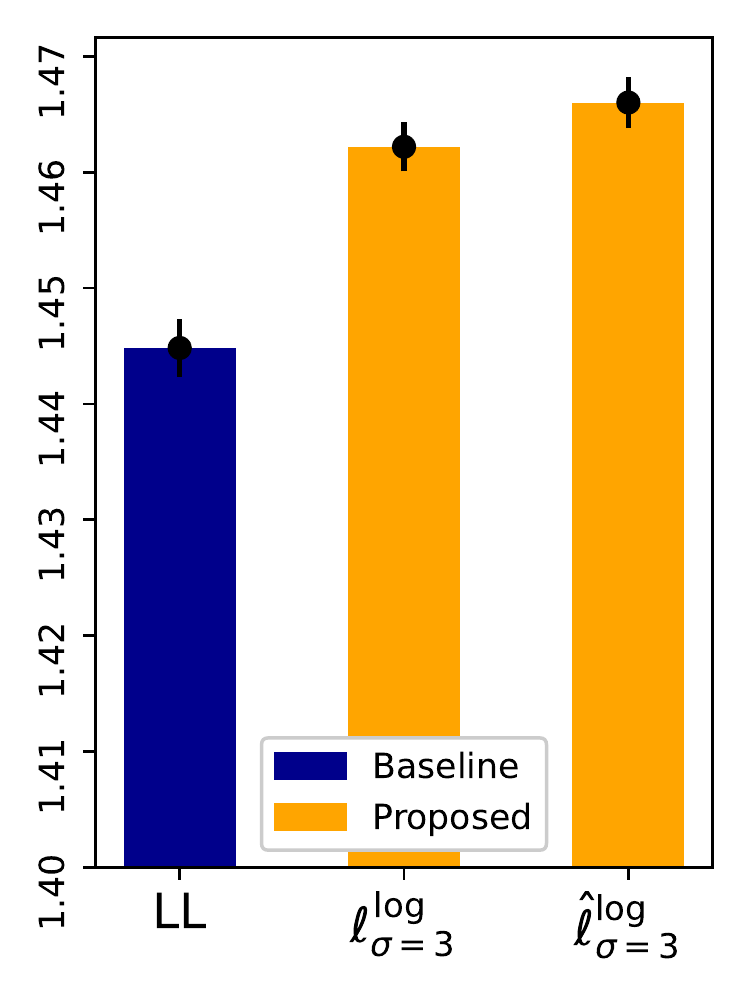}
    \includegraphics[width=0.25\linewidth]{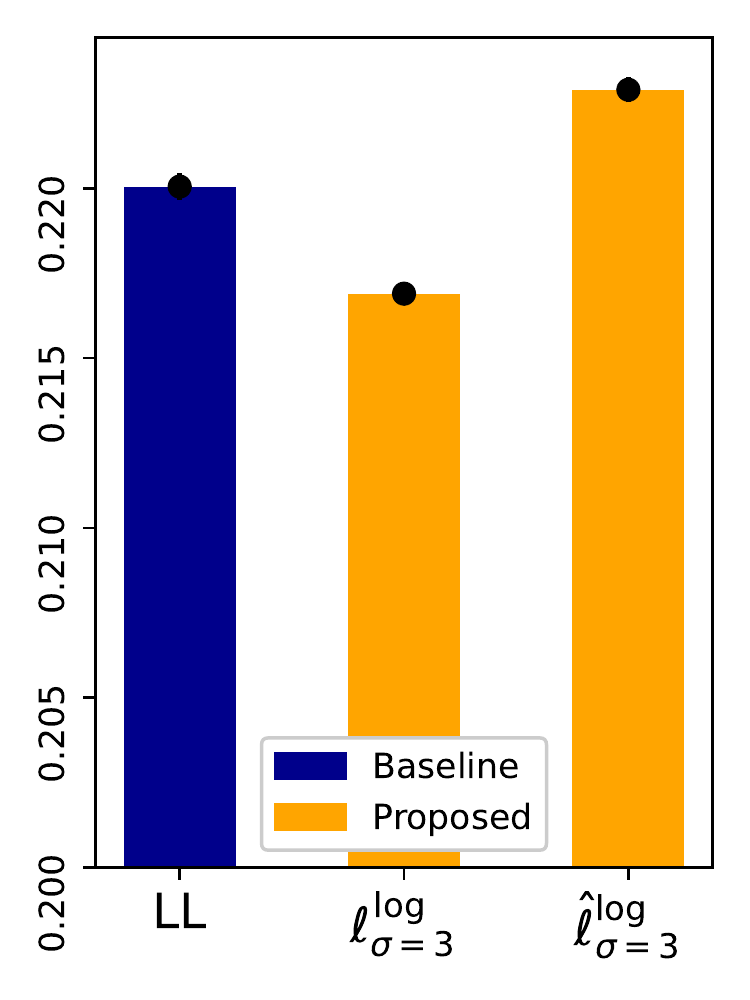}
    \includegraphics[width=0.25\linewidth]{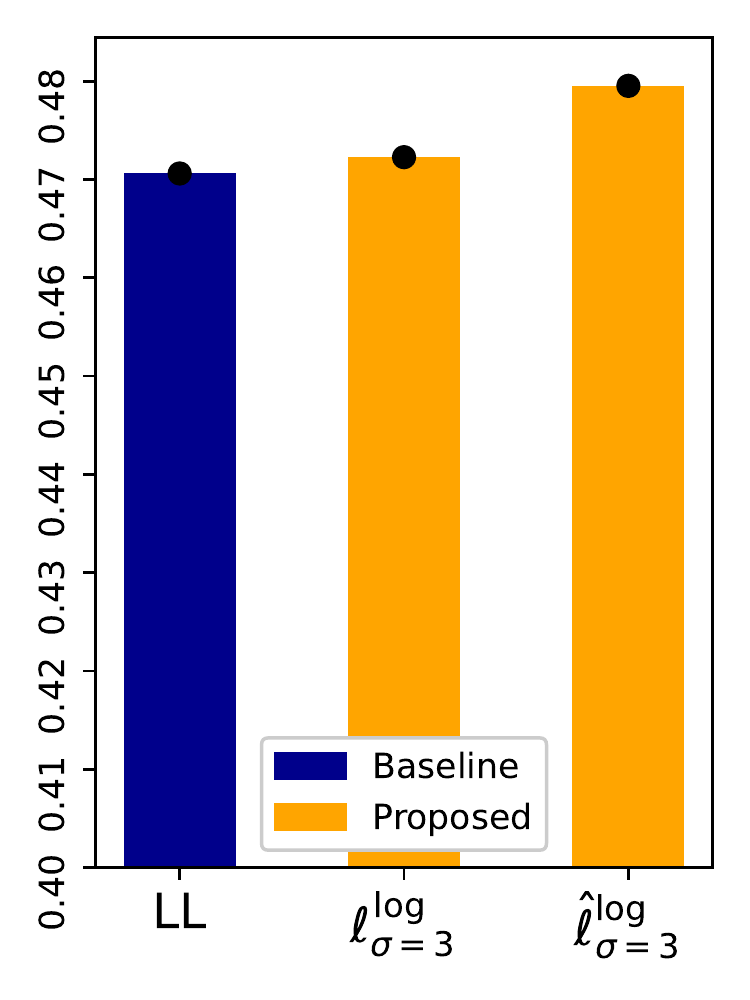}
    \caption{Detailed Performance Metrics for DeepFM. \textbf{Left: Welfare} (higher is better). \textbf{Center: AUC loss} (lower is better). \textbf{Right: Logistic loss} (lower is better). FFor each plot, from left to right: (baselines, blue) logistic loss, (proposed, yellow) $\ell^{\log}_{\sigma = 3}$ (defined in~\eqref{eq:defn_log_surrogate}), and $\hat{\ell}^{\log}_{\sigma}$ (\eqref{eq:defn_log_surrogate} with $y_i$'s replaced by outputs from teacher network).}
    \label{fig:detailed_deepfm}
\end{figure}

\paragraph{Additional Experimental Results.} We plot DeepFM's welfare, AUC loss, and logistic loss in Figure~\ref{fig:detailed_deepfm}.

\begin{figure}[hbt]
    \centering
    \includegraphics[width=0.25\linewidth]{plots/dcn_10_runs.pdf}
    \includegraphics[width=0.25\linewidth]{plots/dcn_10_runs_auc.pdf}
    \includegraphics[width=0.25\linewidth]{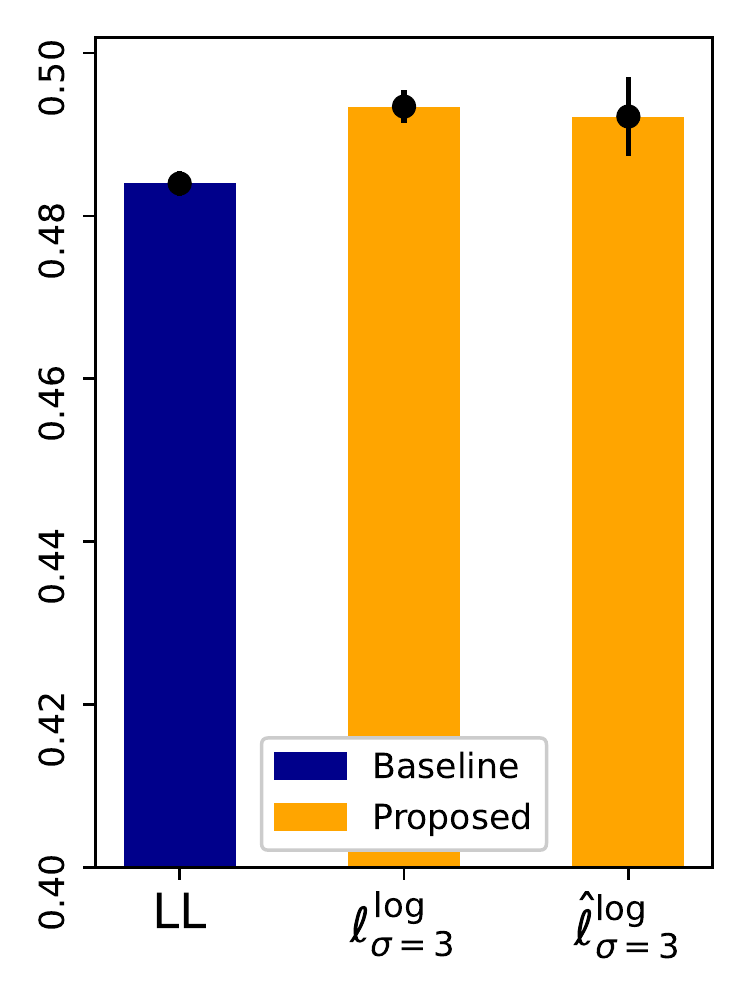}
    \caption{Detailed Performance Metrics for DCN. \textbf{Left: Welfare} (higher is better). \textbf{Center: AUC loss} (lower is better). \textbf{Right: Logistic loss} (lower is better). For each plot, from left to right: (baselines, blue) logistic loss, (proposed, yellow) $\ell^{\log}_{\sigma = 3}$ (defined in~\eqref{eq:defn_log_surrogate}), and $\hat{\ell}^{\log}_{\sigma}$ (\eqref{eq:defn_log_surrogate} with $y_i$'s replaced by outputs from teacher network).}
    \label{fig:detailed_dcn}
\end{figure}

We also plot DCN's welfare, AUC loss, and logistic loss in Figure~\ref{fig:detailed_dcn}.

The exact values of the metrics and the associated standard errors for DeepFM can be found in Table~\ref{tab:deepfm}. The exact values of the metrics and the associated standard errors for DCN can be found in Table~\ref{tab:dcn}.

\begin{table}[ht]
    \centering
    \begin{tabular}{*4c}
        \toprule
        \multicolumn{4}{c}{DeepFM}\\
        \midrule
        {}   & Welfare   & AUC Loss    & Logloss \\
        Logistic Loss   &  $1.4448\pm 0.0025$ & $0.2200\pm 0.0004$   & $0.4706\pm 0.0005$  \\
        $\ell^{\log}_{\sigma = 3}$   &  $1.4622\pm 0.0021$ & $0.2169\pm 0.0003$   & $0.4723\pm 0.0009$  \\
        $\hat{\ell}^{\log}_{\sigma = 3}$   &  $\mathbf{1.4660\pm 0.0022}$  &  $0.2229\pm 0.0004$   & $0.4795\pm 0.0009$ \\
        \bottomrule
    \end{tabular}
    \caption{Results for DeepFM.}
    \label{tab:deepfm}
\end{table}

\begin{table}[ht]
    \centering
    \begin{tabular}{*4c}
        \toprule
        \multicolumn{4}{c}{DCN}\\
        \midrule
        {}   & Welfare   & AUC Loss    & Logloss \\
        Logistic Loss   &  $1.4622\pm 0.0026$ & $0.2173\pm 0.0004$   & $0.4840\pm 0.0015$  \\
        $\ell^{\log}_{\sigma = 3}$   &  $\mathbf{1.4663\pm 0.0028}$ & $0.2146\pm 0.0005$   & $0.4934\pm 0.0020$  \\
        $\hat{\ell}^{\log}_{\sigma = 3}$   &  $1.4650\pm 0.0027$  &  $0.2115\pm 0.0007$   & $0.4922\pm 0.0048$ \\
        \bottomrule
    \end{tabular}
    \caption{Results for DCN.}
    \label{tab:dcn}
\end{table}

As we can see from the results, the proposed methods significantly improve welfare at a minimal cost (if any) to AUC loss. While logistic loss seems to be negatively affected, the impact is relatively small and we conjecture that better tuning the model architecture could resolve the issue.

Finally, we report below the per epoch runtime of the methods. 

\begin{table}[th]
    \centering
    \begin{tabular}{*4c}
        \toprule
        \multicolumn{4}{c}{DCN}\\
        \midrule
        {}   & Logloss   &  Pairwise Loss ($\ell^{\log}_{\sigma = 3}$)    & Pairwise Loss + Student-Teacher ($\hat{\ell}^{\log}_{\sigma = 3}$) \\
        Absolute &  $2635.13\pm 10.86$ & $ 2700.87\pm 8.91$   & $2700.4\pm 6.59$  \\
        Relative   &  100\% & 102.5\%   & 102.5\%\\
        \bottomrule
    \end{tabular}
    \caption{Runtime (in seconds) comparison for DCN. We take the average over 15 epochs and report the standard deviation.}
    \label{tab:deepfm_time}
\end{table}

\begin{table}[th]
    \centering
    \begin{tabular}{*4c}
        \toprule
        \multicolumn{4}{c}{DCN}\\
        \midrule
        {}   & Logloss   &  Pairwise Loss ($\ell^{\log}_{\sigma = 3}$)    & Pairwise Loss + Student-Teacher ($\hat{\ell}^{\log}_{\sigma = 3}$) \\
        Absolute &  $5292.2\pm 57.39$ & $ 5333.67\pm 56.62$   & $5330\pm 48.11$  \\
        Relative   &  100\% & 100.8\%   & 100.7\%\\
        \bottomrule
    \end{tabular}
    \caption{Runtime (in seconds) comparison for DCN. We take the average over 15 epochs and report the standard deviation.}
    \label{tab:dcn_time}
\end{table}

These results show that for the more complex DCN model the added cost of the pairwise losses is relatively negligible. For DeepFM, the added computation cost is around 2.5\% (60 seconds), which is a reasonable price to pay for significantly improved welfare performance. It is possible that our implementation of the loss is not the most efficient, and additional optimizations may further decrease the overhead.

\end{document}